\documentclass[12pt]{article}
\usepackage{amssymb}
\SetSymbolFont{largesymbols}{normal}{OMX}{lcmex}{m}{n}
  
\catcode`@=11
\def\secteqno{\@addtoreset{equation}{section}%
\def\theequation{\thesection.\arabic{equation}}}
\catcode`@=12
\secteqno
\def\dd{\hbox{\,\Large$\triangleright$}}
\newcommand{\be}{\begin{equation}}
\newcommand{\ee}{\end{equation}}
\newcommand{\bea}{\begin{eqnarray}}
\newcommand{\eea}{\end{eqnarray}}
\newcommand{\bref}[1]{(\ref{#1})}
\newcommand{\nn}{\nonumber}

\def\dig#1{\setbox0=\hbox{$#1M$}
	\hskip.06\wd0 \vrule width.08\wd0 height.63\wd0 depth.01\wd0 
	\vrule width.37\wd0 height.63\wd0 depth-.55\wd0 \hskip-.4\wd0
	\vrule width.25\wd0 height.36\wd0 depth-.28\wd0 
	\vrule width.08\wd0 height.36\wd0 depth-.17\wd0 \hskip.14\wd0}
\def\digamma{{\mathpalette\dig{}}}
\def\bop#1{\setbox0=\hbox{$#1M$}\mkern1.5mu
	\vbox{\hrule height0pt depth.1\ht0
	\hbox{\vrule width.1\ht0 height.8\ht0 \kern.8\ht0
	\vrule width.1\ht0}\hrule height.1\ht0}\mkern1.5mu}
\def\bo{{\mathpalette\bop{}}} 		

\font\big=cmmi12 scaled 1440
\def\dd{\hbox{\big\char'056}}		

\def\don#1#2{{\buildrel{\mkern2.5mu\raise-.1em\hbox{$\scriptstyle#1$}\mkern-2.5mu}\over{#2}}}	
\def\dron#1#2{{\buildrel{{\raise-.1em\hbox{$\scriptstyle#1$}}}\over{#2}}}		

\begin{document}
\begin{flushright}
\parbox{4.2cm}
{2017,~January 23\\
KEK-TH-1951 \hfill \\YITP-SB-17-3\hfill \\
}
\end{flushright}

\vspace*{1.1cm}

\begin{center}
 {\Large\bf Manifestly T-dual formulation of AdS space }\\
\end{center}
\vspace*{1.5cm}
\centerline{\large Machiko Hatsuda$^{\dagger \natural }$\footnote{mhatsuda@post.kek.jp, mhatsuda@juntendo.ac.jp
}, Kiyoshi Kamimura$^{\dagger }$\footnote{kamimura@ph.sci.toho-u.ac.jp}
and Warren Siegel$^\star$
\footnote{siegel@insti.physics.sunysb.edu
,
http://insti.physics.sunysb.edu/{\tt \~{}}siegel/plan.html}
}
\begin{center}
$^{\dagger}$\emph{Physics Division, Faculty of Medicine,
 Juntendo University, Chiba 270-1695, Japan}
\\
$^{\natural}$\emph{KEK Theory Center, High Energy Accelerator Research 
Organization,\\
Tsukuba, Ibaraki 305-0801, Japan} 
\\
$^\star$\emph{C. N. Yang Institute for Theoretical Physics
State University of New York, Stony Brook, NY 11794-3840}
\vspace*{0.5cm}
\\

\end{center}

\vspace*{1cm}

\centerline{\bf Abstract}
\vspace*{0.5cm}

We present a manifestly T-dual formulation of 
curved spaces such as an AdS space.
For group manifolds related by the orthogonal vielbein fields
the three form $H=dB$ in the doubled space
is universal at least locally.
We construct an affine nondegenerate doubled bosonic AdS algebra to define 
the AdS space with the   Ramond-Ramond flux.
The non-zero commutator of the left and right momenta
leads to that the left momentum is in an AdS space 
while the right momentum is in a dS space.  
Dimensional reduction constraints and the physical AdS algebra are 
shown to preserve all the doubled coordinates.
\vfill 

\thispagestyle{empty}
\setcounter{page}{0}
\newpage

\section{Introduction and conclusions}

T-duality is  one of the most characteristic features of string theories.
The T-duality symmetry exists in its low energy effective theory
described by the massless modes.
Such a stringy gravity theory is a theory of the gravitational field $G_{mn}$, the $B_{mn}$ field and the dilaton field.
The general coordinate transformation 
is generalized in a T-duality covariant way. 
It is shown to be generated by the zero mode of 
the affine nondegenerate doubled Lie algebra \cite{Siegel:1993xq}.
This manifestly T-dual formulation is the procedure to construct gravity theories and  
it is  being 
developed in \cite{Siegel:1994xr}-
 \cite{Hatsuda:2015cia}.
The procedure contains roughly  two steps:
doubling the d-dimensional coordinates to manifest the O(d,d) T-duality symmetry and 
imposing constraints to reduce 
a half of the doubled coordinates
preserving the T-duality symmetry.  
For a flat space the procedure is straightforward,
however for curved spaces it becomes nontrivial.

T-duality along a non-abelian isometry had been proposed 
 \cite{delaOssa:1992vci} and non-abelian T-duality in AdS spaces  
has been investigated  in for example \cite{Klimcik:2002zj,
Sfetsos:2010uq},
in which there remain many interesting problems to solve.
Recently the equivalence between the integrable deformation of the AdS superstring
 and the non-abelian T-duality was proposed in \cite{Hoare:2016wsk} and 
has been developed  in \cite{Borsato:2016pas}.
It was shown that  the nonlocal charges of a string 
are equal to the Noether charges of a string in the T-dualized space
for a flat space and the pp-wave space \cite{Hatsuda:2006ts}
which gives a relation between the integrability and  the abelian T-duality.
The superstring in the AdS$_5\times$S$^5$ space has integrability  \cite{Bena:2003wd}, 
and the nonlocal charges generate the Yangian algebra
as shown in \cite{Hatsuda:2005te}
based on the Hamiltonian formulation of the AdS string \cite{Hatsuda:2001xf}.
In order to clarify the features of the non-abelian T-duality
and its integrability
 the manifestly T-dual formulation of AdS space will be useful
as the doubled space analysis.
The superspace approach to the AdS space with manifestly T-duality is presented in \cite{Polacek:2016nry}
 based on the super-AdS algebra in 
\cite{Hatsuda:2014qqa}.

The double field theory on group manifolds has been studied in \cite{Blumenhagen:2014gva}. 
Our formulation is different from theirs in the following points, though
it has some overlap:
\begin{itemize}
  \item {Vielbein fields are used to express the scalar curvature of the stringy gravity theory
rather than the generalized metric.
This is necessary to couple spacetime fermions.}
\item{
First class constraints are used to reduce a half of the doubled coordinates 
 rather than 
the gauge fixing condition for the section constraint.
This preserves the T-duality covariant local symmetry. }
\item{
The Wess-Zumino term in the worldsheet action 
is also doubled in order to give
both the covariant derivative currents and the symmetry generator currents 
for both the left and right modes.
The usual Wess-Zumino-Witten procedure gives
the covariant derivative current for the right mode
and the symmetry generator current for the left mode.
This is necessary to describe N=2 supersymmetric theories.
}
\end{itemize}

The manifestly T-dual formulation is 
a gauge theoretical formulation of the stringy gravity:
The gauge theory is a theory of the gauge field $A_m{}^I$ with
a gauge group generated by $G_I$.
The covariant derivative $\nabla_m$ is an essential operator which gives
the gauge transformation rule, and
its commutator gives the field strength $F_{mn}{}^I$
\bea
&p_m=\frac{1}{i}\partial_m\to\nabla_m=p_m+A_m{}^IG_I&\nn\\
&\delta_\lambda \nabla_m=i\lbrack \Lambda,\nabla_m]~,~\Lambda=\lambda^IG_I
\Rightarrow
\delta_\lambda A_m=-\partial_m\lambda&\nn\\&
~i\lbrack \nabla_m,\nabla_n]=F_{mn}{}^IG_I~~~.&
\eea 
The gravity theory is a gauge theory of the vielbein $e_a{}^m$ \cite{Siegel:1999ew}.
The covariant derivative $\nabla_a$ gives  
the general coordinate transformation rule as
\bea
&p_m\to\nabla_a=e_a{}^mp_m&\nn\\&\delta_\lambda\nabla_a=
i\lbrack \Lambda,\nabla_a]~,~\Lambda=\lambda^mp_m\Rightarrow\delta_\lambda e_a{}^m={\cal L}_\lambda e_a{}^m=\lambda^n\partial_ne_a{}^m
-e_a{}^n\partial_n\lambda^m~.&
\eea
In order to obtain the curvature tensor 
the Lorentz generator $s_{mn}$ must be taken into account as the generator.
The covariant derivative includes the Lorentz connection $\omega_{a}{}^{mn}$ and
its commutator gives
the Riemannian curvature tensor $R_{ab}{}^{cd}$ as
\bea
& \nabla_a=e_a{}^mp_m+\frac{1}{2}\omega_{a}{}^{mn}s_{mn}~,~
i\lbrack\nabla_a,\nabla_b]=T_{ab}{}^c\nabla_c+\frac{1}{2}R_{ab}{}^{cd}s_{cd}~~~,
&
\eea
where the torsion constraint $T_{ab}{}^c=0$ 
relates  $e_a{}^m$ and  $\omega_{a}{}^{mn}$. 
This is extended to the stringy gravity by doubling the momentum 
including the winding mode as
 $p_m\to \left(p_{\rm m}(\sigma),\partial_\sigma x^{\rm m}(\sigma)\right)$ \cite{Siegel:1993xq}.
The stringy gravity is a gauge theory of the doubled vielbein field 
$e_{\underline{\rm a}}{}^{\underline{\rm m}}\in$O(d,d)/O(d-1,1)$^2$ 
with the doubled indices $_{\underline{\rm m}}
=(_{\rm m},~^{\rm m})$ and $_{\underline{\rm a}}$.
 $e_{\underline{\rm a}}{}^{\underline{\rm m}}$ is considered as a function of $x^{\rm m}$ here,
 although it will be considered as a function of the doubled coordinates later.
The stringy covariant derivative $\dd_{\underline{\rm a}}(\sigma)$ gives  
the gauge transformation rule $\delta_\lambda e_{\underline{\rm a}}{}^{\underline{\rm m}}$ as
\bea
&p_m\to\nabla_a\to
\dd_{\underline{\rm a}}(\sigma)=e_{\underline{\rm a}}{}^{\rm m}p_{\rm m}+
e_{\underline{\rm a}}{}_{\rm m}\partial_\sigma x^{\rm m}&\nn\\
&\delta_\lambda\dd_{\underline{\rm a}}(\sigma)
=i\lbrack \Lambda,\dd_{\underline{\rm a}}(\sigma)]~,~
\Lambda=\int d\sigma\left(
\lambda^{\rm m}p_{\rm m}+\lambda_{\rm m}\partial_\sigma x^{\rm m}\right)~~.&
\eea
If $e_{\underline{\rm a}}{}^{\underline{\rm m}}$ is written in terms of $G_{mn}$ and $B_{mn}$ 
with the doubled Minkowski metric $\hat{\eta}^{\underline{\rm ab}}$,
then the T-duality covariantized general coordinate transformation is given by
\bea
e_{\underline{\rm a}}{}^{\underline{\rm m}}
\hat{\eta}^{\underline{\rm  ab}}e_{\underline{\rm  b}}{}^{\underline{\rm n}}
={\renewcommand{\arraystretch}{1.2}
\left(\begin{array}{cc}G^{mn}&G^{ ml}
B_{ln}\\
-B_{ ml}G^{ ln}&
G_{ mn}-B_{ ml}G^{ lk}B_{kn}
\end{array}
\right)}\Rightarrow 
{\renewcommand{\arraystretch}{1.2}
\left\{\begin{array}{l}
\delta_\lambda G_{mn}={\cal L}_\lambda G_{mn}\\
\delta_\lambda B_{mn}={\cal L}_\lambda B_{mn}-
\partial_{[m}\lambda_{n]}
\end{array}\right.}
\eea
The curvature tensors are obtained by taking into account the
Lorentz generator.
The left and right Lorentz generators  
 $S_{\underline{mn}}=(S_{mn},~S_{{m'n'}})$ are introduced
in the left and right basis, $P_{\underline{m}}=(P_m,~P_{m'})$ wtih
$P_m=(p_{\rm m}+\partial_\sigma x^{\rm m})/\sqrt{2},~
P_{m'}=(p_{\rm m}-\partial_\sigma x^{\rm m})/\sqrt{2}$
 in the unitary gauge.
The consistency of the ``affine" Lie algebra requires the nondegeneracy of the group metric. The resultant algebra is
the affine nondegenerate doubled Poincar\'{e} algebra generated by
$
\dron{\scriptscriptstyle\bo}{\dd}_{\underline{M}}(\sigma)=(
S_{\underline{mn}},~P_{\underline{m}},~\Sigma^{\underline{mn}})$
where $\Sigma^{\underline{mn}}$ is necessary for the nondegeneracy.
The stringy covariant derivative $\dd_{\underline{A}}$ includes the doubled vielbein 
$E_{\underline{A}}{}^{\underline{M}}$, and its commutator  
becomes
\bea
& \dd_{\underline{A}}=E_{\underline{A}}{}^{\underline{M}}\dron{\scriptscriptstyle\bo}{\dd}_{\underline{M}}~,~
i\lbrack\dd_{\underline{A}},\dd_{\underline{B}}]=T_{\underline{AB}}{}^{\underline{C}}\dd_{\underline{C}}&
\eea
where the curvature tensors  are included in the torsions as
 $T_{\underline{PP}}{}^{\underline{S}}$.

  In this paper we extend this manifestly T-dual formulation
in the asymptotically flat space  \cite{
Siegel:1993xq}, \cite{Siegel:2011sy}-
  \cite{Hatsuda:2015cia} to 
curved spaces such as an AdS space. 
It is interesting and important to
discuss supersymmetric case. 
However we restrict bosonic algebra in this paper leaving the supersymmetric case 
for future works. 
Our main result is the affine nondegenerate doubled bosonic AdS algebra
\bref{result1}-\bref{2AdSsym}
which defines the AdS space with manifest T-duality and 
generates the T-duality covariant general coordinate transformation.

The results are based on the following observations which we found 
in this paper:
\begin{itemize}
 \item {
\begin{description}
  \item[Local universality of the three form $H=dB$ in the doubled space]
\end{description}
 For curved spaces described by Lie algebras
 the three form $H=dB$ in the doubled space is universal
 at least locally. 
 The doubled space three form of a group manifold is given by 
$H=\frac{1}{3!}J^{\underline{I}}\wedge
J^{\underline{J}}\wedge J^{\underline{K}}f_{\underline{IJK}}$ 
with the left-invariant current $J^{\underline{I}}$,
 the structure constant $f_{\underline{IJK}}
=f_{\underline{IJ}}{}^{\underline{L}}\eta_{\underline{LK}}$
 and 
the nondegenerate group metric $\eta_{\underline{IJ}}$.
Doubled indices run over the left and right indices
$_{\underline{I}}=(_{I,~I'})$.
The doubled space three form $H$ is the sum of fluxes chained by  
T-duality, $H_{\rm  IJK}\to f_{\rm IJ}{}^{\rm K}\to Q_{\rm I}{}^{\rm JK}\to R^{\rm IJK}$  
in the base of $(p_{\rm m},\partial_\sigma x^{\rm m})$
as $_{\underline{I}}=(_{\rm I},~{}^{\rm I})$. 
The flat space operators are denoted with ``$\don{\scriptscriptstyle\bo}{~}$~".
The stringy covariant derivatives in curved and flat spaces are 
by the vielbein $E_{\underline{I}}{}^{\underline{M}}$. 
We show that the doubled space three form $H$ in curved and flat spaces 
are the same 
\bea
&&
\dd_{\underline{I}}=E_{\underline{I}}{}^{\underline{M}}
\dron{\scriptscriptstyle\bo}{\dd}_{\underline{M}}~,~
E_{\underline{I}}{}^{\underline{M}}E_{\underline{J}}{}^{\underline{N}}
\eta_{\underline{MN}}=\eta_{\underline{IJ}}~\to
H=\dron{\scriptscriptstyle\bo}{H}~~~.
\eea
The gauge transformation of $B$ field is also 
recognized as a T-duality rotation.
The dilaton factor may play a  role for a
different value of the doubled space three form.
\par\vskip 6mm
The doubled flat space is described by
the cosest group G/H
 where G is the nondegenerate doubled Poincar\'{e} group and 
 H is the doubled Lorentz group $\times$ (dimensional reduction constraint) \cite{Hatsuda:2014aza}.
The doubled space three form $\don{\scriptscriptstyle\bo}{H}=d\don{\scriptscriptstyle\bo}{B}$ in the flat space
 belongs to a trivial
class of the Chevalley-Eilenberg (CE) cohomology \cite{DeAzcarraga:1989vh}
 of the cosest G/H.
We have shown that  $\don{\scriptscriptstyle\bo}{B}_{\underline{IJ}}$ in the doubled flat space 
 is constant, 
so the Wess-Zumino term  $\don{\scriptscriptstyle\bo}{B}=\don{\scriptscriptstyle\bo}{B}_{\underline{IJ}}\don{\scriptscriptstyle\bo}{J}{}^{\underline{I}}\wedge \don{\scriptscriptstyle\bo}{J}{}^{\underline{J}}$ is bilinears of the left-invariatn currents \cite{Hatsuda:2014aza,Hatsuda:2015cia}. 
For the  nondegenerate doubled AdS coset group,
the three form
 $\don{\circ}{H}$  is closed,   $d\don{\circ}{H}=0$,
 but it belongs to a nontrivial class of the CE cohomology.
The supersymmetry will change the situation 
 as the supr-AdS group in the non-doubled space \cite{Hatsuda:2004vi}. 

}

 \item{
\begin{description}
  \item[ Spontaneous symmetry breaking by the Ramond-Ramond  flux
] 
\end{description}

When the Ramond-Ramond (RR)  flux has a non-zero vacuum expectation value,
$\langle 0|F_{\rm RR}^{\alpha\beta'}|0\rangle\neq 0$,
the Lorentz symmetry is broken;
the full Lorentz symmetry is broken into its subgroup
and the left and right Lorentz symmetries in the doubled space 
are broken into a linear combination of them. 
It is natural to expect non-zero commutator of the left and right 
 momenta $p_a$ and $p_{a'}$ 
 as well as the non-zero anticommutator of the left and right supercovariant derivatives.
We found that the nondegenerate doubled bosonic  AdS algebra includes
\bea
&&\lbrack p_{a},p_{b}]=i(\frac{1}{r_{\rm AdS}{}^2}s_{ab}+\sigma_{ab})~,~
\lbrack p_{a'},p_{b'}]=i(\frac{1}{r_{\rm AdS}{}^2}s_{a'b'}+\sigma_{a'b'})\nn\\
&&\lbrack p_{a},p_{b'}]=i(\frac{1}{r_{\rm AdS}{}^2}s_{ab'}+\sigma_{ab'})
\label{AdSppp}
\eea
where $s_{ab}$'s and $\sigma^{ab}$'s are nondegenerate partners. 
$r_{\rm AdS}$ is the AdS radius.
The momentum $p_a$ is a d-dimensional vector and 
the doubled momentum must be a SO(d,d) vector.
Therefore the third equation in \bref{AdSppp} leads to that the left and right momenta are embedded in  SO(d,d+1). 
The left moving momentum is in an AdS space while  
the right moving momentum is forced to be in a  dS space. 
 This phenomena is similar to the point discussed in
\cite{Klimcik:2002zj}.
Now the doubled Lorentz group  is SO(d,d) instead of SO(d$-$1,1)$\times$SO(1,d$-$1).
Similarly the d-dimensional sphere is described by the coset
SO(d+1,d)/SO(d,d) in the doubled space.
\par\vskip 6mm

The RR flux of the AdS$_5\times$S$^5$ in the type IIB superstring theory
breaks the SO(9,1) Lorentz symmetry  into  
SO(4,1)$\times$SO(5).
Naive doubling of the Lorentz subgroup 
does not give the correct number of  degrees of freedom of $G_{mn}$ and $B_{mn}$. 
The number of dimensions of the naive coset, 
 O(10,10)/[SO(4,1)$\times$SO(1,4)$\times$SO(5)$^2$ ]
 is not $10^2$. 
We solve this puzzle; 
now the doubled Lorentz group is O(5,5)$^2$
so the coset becomes   O(10,10)/SO(5,5)$^2$ whose 
number of dimensions 
coincides with the number of degrees of freedom of $G_{mn}$ and $B_{mn}$.
}
\item{
\begin{description}
  \item[Nondegenerate non-abelian group]
\end{description}
A general method to construct a 
nondegenerate group is the followings:   
Copy the subgroup H$_0$ of a coset group G/H$_0$ to H$_1$
and take the direct  product: G$\to$G$\times$H$_1$ \cite{Bonanos:2009wy}.
Make subgroups by the semidirect product of H and $\check{\rm H}$
from H$_0$ and H$_1$, H$_0\times$H$_1 \to$ H $\ltimes \check{\rm H}$,
where  H is generated by the vector type currents 
and $\check{\rm H}$ is generated by the axial vector type currents.
Then the nondegenerate group metric for H and $\check{\rm H}$
is introduced as
\bea
{\rm G}&\to&{\rm G}\times{\rm H}_1\label{ndhhc} \\
{\rm H}_0&\to&{\rm H}_0\times{\rm H}_1\to {\rm H}\ltimes\check{\rm H},~{\rm with}~
{\rm tr}(h\check{h})=\eta_{h\check{h}},~h,\check{h}
\in{\rm Lie~algebras~of}~{\rm H},\check{\rm H}~~.\nn
\eea
H and $\check{\rm H}$ correspond to the Lorentz group and its nondegenerate partner.
}
\item{
\begin{description}
  \item[Dimensional reduction constraints for nondegenerate partners] 
\end{description}
These dimensions of nondegenerate partners are unphysical and reduced by 
imposing dimensional reduction constraints.
For an element $a$ of a group A 
the covariant derivative and the symmetry generator
are calculated from $a^{-1}da$ and   $(da)a^{-1}$ respectively. 
We denote a group A which is generated by the covariant derivative
 and $\tilde{\rm A}$ which is generated by the symmetry generator.
A$\times\tilde{\rm A}$ acts on $a$ by $a$$\to\tilde{c}ab^{-1}$,~
$b\in$A and $\tilde{c}\in\tilde{\rm A}$. 
The nondegenerate coset group is obtained from
 \bref{ndhhc} 
as G/H$_0$ $\to$G$\times$H$_1$/H$\ltimes$$\check{\rm H}$.
However H and $\check{\rm H}$ can not be imposed as 
first class constarints because of the Schwinger term 
for the nondegeneracy.
Instead H and $\tilde{\check{\rm H}}$ can be imposed as
first class constraints, since the covariant derivative and the symmetry generator commute.
So the obtained coset is
\bea
\frac{\rm G}{{\rm H}_0}\stackrel{\rm nondegenerate}{\longrightarrow}
\frac{{\rm G}\times {\rm H}_1}
{ {\rm H}\times \tilde{\check{\rm H}}}
 \eea
The d-dimensional AdS space is described in the doubled space 
with nondegeneracy as ;
\bea
 \frac{{\rm SO(d-1,2)}}{{\rm SO(d-1,1)}}\stackrel{\rm double}{\longrightarrow}
 \frac{{\rm SO(d,d+1)}}{{\rm SO(d,d)}_0}\stackrel{\rm nondegenerate}{\longrightarrow}
 \frac{{\rm SO(d,d+1)}\times{\rm SO(d,d)}_1}{
 {\rm SO(d,d)}\times\tilde{\check{\rm SO}}{\rm (d,d)}}
 \eea 
 Similarly the d-dimensional sphere is described in the doubled space 
with nondegeneracy as;
\bea
 \frac{{\rm SO(d+1)}}{{\rm SO(d)}}
 \stackrel{\rm double}{\longrightarrow}
 \frac{{\rm SO(d+1,d)}}{{\rm SO(d,d)}_0}\stackrel{\rm nondegenerate}{\longrightarrow}
 \frac{{\rm SO(d+1,d)}\times{\rm SO(d,d)}_1}{
 {\rm SO(d,d)}\times\tilde{\check{\rm SO}}{\rm (d,d)}}
 \eea
 For a special case of AdS$_5\times$S$^5$
 we find that the group structure of the bosonic part is 
 \bea
 \frac{{\rm SO(5,6)}\times{\rm SO(5,5)}_1}{
 {\rm SO(5,5)}\times\tilde{\check{\rm SO}}{\rm (5,5)}}
  \times \frac{{\rm SO(6,5)}\times{\rm SO(5,5)}_1}{
 {\rm SO(5,5)}\times\tilde{\check{\rm SO}}{\rm (5,5)}}~~~.
 \eea  
}
\item{
\begin{description}
  \item[Dimensional reduction constraints for doubled momenta] 
\end{description}
A half of the doubled momenta is reduced by 
the dimensional reduction constraint, 
$\phi_a=\tilde{P}_a-\tilde{P}_{a'}\delta_a{}^{a'}=0$.
The symmetry generators of the affine algebras are
 $\tilde{P}$ for momentum and $\tilde{S}$ for Lorentz generator. 
The physical AdS algebra is genereted by the physical momentum $\tilde{P}_{{\rm total};a}$ and the physical Lorentz generator $\tilde{S}_{{\rm total};ab}$
without gauge fixing for a half coordinates;
\bea
&\tilde{P}_{{\rm total};a}=\tilde{P}_a+\tilde{P}_{a'}\delta_a{}^{a'}+\cdots~,~\tilde{S}_{{\rm total};ab}=\tilde{S}_{ab}-
\tilde{S}_{a'b'}\delta_a{}^{a'}\delta_b{}^{b'}+\cdots~&\nn\\
&\lbrack \int \tilde{P}_{{\rm total};a},\int \tilde{P}_{{\rm total};b}]=\frac{i}{r_{\rm AdS}{}^2}\int \tilde{S}_{{\rm total};ab}
&\label{phyAdS}
\eea
where $\cdots$ includes first class constraints 
and the left-right mixing term.
The equation \bref{phyAdS} is the expected physical AdS algebra.
}
\end{itemize}
 
The organization of the paper is the following.
In section 2 we 
explain the procedure of the manifestly T-dual formulation.
Notations are listed there.
The general method to construct a nondegenerate  Lie algebra and  
to double the Lie group is presented.  
Then affine extension of the obtained Lie algebra is performed.
The equation on the $B$ field is obtained.
The  computation of the zero mode of the affine Lie algebra is demonstrated. 
In section 3 the manifestly T-dual formulation of the flat space is reviewed.
The $B$ field is constant where the dilatation operator plays a role.
The relation between the dimensional reduction constraints and the section condition is explained.
 In section 4 the manifestly T-dual formulation of curved spaces is presented.
After examining the relation between the flat covaiant derivative and the  curved space covariant derivatives of  group manifolds,
 the $B$ field and the three form $H=dB$ are obtained.
  In section 5 the manifestly T-dual formulation of an AdS space is presented.
It is explained that the RR flux naturally gives
the left and right mixing Lorentz generators.
The nondegenerate doubled AdS algebra is obtained,
then affine extension is performed.
The dimensional reduction constraints and the physical AdS algebra are 
obtained with manifest T-duality.

\par
\vskip 6mm

\section{ Manifestly T-dual formulation}

At first we 
explain the procedure of the manifestly T-dual formulation. 
List of notations is also in subsection 2.1.
In subsection 2.2.1 the general method to construct a nondegenerate Lie algebra is presented.
In subsection 2.2.2
 it is shown that doubled coordinates are convenient to 
describe the closed string mechanics and  doubling the whole group
gives simpler treatment of the system. 
In subsection 2.3
 we extend the obtained nondegenerate doubled Lie algebra to
 affine Lie algebras generated by the string covariant derivative $\dd_I$
 and the string symmetry generator $\tilde{\dd}_I$.
 The $B$ field appears in the string covaiant derivatives $\dd_I$
 as the relative coefficient of the particle covariant derivative $\nabla_I$
 and the $\sigma$ component of the left-invariant current $J_1{}^I$.
 The affine Lie algebra gives the equation on the $B$ field.
 The space with manifest T-duality is defined by the affine Lie algebra
 generated by the string covariant derivative.
 The gauge symmetry of the space is generated by the affine Lie derivative.
The  computation of the zero mode of the affine Lie algebra is demonstrated. 

\par
\vskip 6mm
\subsection{Procedure and notations}

In this subsection we present the manifestly T-dual formulation 
and notations 
proposed in
\cite{Hatsuda:2014qqa}-
\cite{Hatsuda:2015cia} based on \cite{Siegel:1993xq}-
\cite{Polacek:2013nla
}.
The procedure is the following:
\begin{enumerate}
\item{Extend a Lie algebra to an affine doubled algebra.

Begin with a Lie algebra and extend it in such a way that 
the nondegenerate group metric can be defined 
in order to construct an affine Lie algebra consistently.
Double the whole algebra in order to make T-duality symmetry manifest.
Perform affine extension of the Lie algebra which include 
the nondegenerate group metric as the coefficient of  the  Schwinger
term. }

\item {Construct the covariant derivative and the symmetry generator
for a string action with manifestly T-duality.

There are two kinds of affine Lie algebras generated by
the covariant derivative $\dd_I$ and the symmetry generator $\tilde{\dd}_I$.
The covariant derivative defines the space which has the T-duality covariant diffeomorphism.
The symmetry generator makes dimensional reduction constraints
 and the physical symmetry algebra.}

\item {Make the curved space  covariant derivative  for a gravity theory with manifestly T-duality.

The covariant derivative in a curved space is obtained by 
multiplying the vielbein field $E_{A}{}^{I}$
on the asymptotic space covariant derivative  $\dd_I$ as
$\dd_A=E_{A}{}^{I}\dd_I$.
The commutator of the curved space covariant deriatives gives 
the torsion. Curvature tensors are included in torsions in this formalism. }

\item{Reduce unphysical dimensions.

A half of the doubled coordinates is reduced by dimensional reduction constraint. The auxiliary dimensions introduced for the nondegeneracy
are also reduced by the dimensional reduction constraints. 
Since dimensional reduction constraints are written in terms of the symmetry generators, the local structure determined by the covariant derivative
is still preserved so the T-duality is manifest.
}
\end{enumerate}

Notations of covariant derivatives and symmetry generators are summarized
as below.
\bea
&&{\rm Covariant ~derivatives:}\nn\\
&&{\renewcommand{\arraystretch}{0.6}
\begin{array}{cccc}
{\rm space}&\begin{array}{c}{\rm Lie~algebra}\\{\rm structure~const.}~
({\rm torsion})
\end{array}
\to&~~~~~{\rm particle}~~~~~\to&~~~~~{\rm string}~~~~~\\\\
&G_I,~f_{IJK}&\nabla_I&\dd_I\\\\
{\rm Poincar\acute{e}}&G_M,~\don{\scriptscriptstyle\bo}{f}_{MNL}&
\dron{\scriptscriptstyle\bo}{\nabla}_M&\dron{\scriptscriptstyle\bo}{\dd}_M\\
\downarrow &&&\\
{\rm Curved}&~~~~~~({T}_{ABC})
&{\nabla}_A=E_A{}^M\dron{\scriptscriptstyle\bo}{\nabla}_M&
{\dd}_A=E_A{}^M\dron{\scriptscriptstyle\bo}{\dd}_M\\\\
{\rm AdS}&G_A,~\don\circ{f}_{ABC}&\dron\circ{\nabla}_A&
\dron\circ{\dd}_A\\
\downarrow &&&\\
{\rm Curved}&~~~~~~({T}_{MNL})
&{\nabla}_M=E_M{}^A\dron\circ{\nabla}_A&
{\dd}_M=E_M{}^A\dron\circ{\dd}_A
\end{array}}
\eea
In curved backgrounds covariant derivatives couple to
gravitational fields, $E_A{}^I$, and 
the commutator of the covariant derivatives gives torsions, ${T}_{IJK}$.
The factorization of the vielbein, $\dd_A=E_A{}^I\dd_I$,
is a general feature of a string theory explained in section 2.2.2.
\bea
&&{\rm Symmetry~ generators:}\nn\\
&&{\renewcommand{\arraystretch}{0.6}
\begin{array}{cccc}
{\rm space}&\begin{array}{c}{\rm Lie~algebra}\\{\rm structure~constant}
\end{array}
\to&~~~~~{\rm particle}~~~~~\to&~~~~~{\rm string}~~~~~\\\\
&G_I,~f_{IJK}&\tilde{\nabla}_I&\tilde{\dd}_I\\\\
{\rm Poincar\acute{e}}&G_M,~\don{\scriptscriptstyle\bo}{f}_{MNL}&
\dron{\scriptscriptstyle\bo}{\tilde{\nabla}}_M&\dron{\scriptscriptstyle\bo}{\tilde{\dd}}_{M}\\
\downarrow &&&\\
{\rm Curved}&&-&-\\\\
{\rm AdS}&G_A,~\don\circ{f}_{ABC}&\dron\circ{\tilde{\nabla}}_A&
\dron\circ{\tilde{\dd}}_A\\
\downarrow &&&\\
{\rm Curved}&&-&-
\end{array}}
\eea
In curved backgrounds 
symmetry generators do not have to generate any global symmetry algebra in general.

\par
\vskip 6mm
\subsection{Nondegenerate doubled Lie algebra }
For affine extension of a Lie algebra the consistency requires
the existence of the nondegenerate group  metric $\eta_{IJ}$ and the 
totally antisymmetric
structure constant with lowered indices $f_{IJK}=f_{IJ}{}^{L}\eta_{LK}=f_{[IJK]}/3!$.
 In subsection 2.2.1
we present a general method to construct a nondegenerate non-abelian group. 
In subsection 2.2.2
after reviewing the string sigma model we double the
 whole group in order to construct both the covariant derivatives and the symmetry generators for both the left and right modes.

\subsubsection{Nondegenerate Lie algebra}

We consider the space governed by the affine Lie algebra.
The consistency of the affine Lie algebra requires the existence of the nondegenerate group metric in the space. 
This nondegenerate group  metric is different from  
 the Killing metric of the Lorentz group.
 The nondegenerate group metric is used to define the 
$\sigma$-diffeomorphism generator in the string worldsheet ${\cal H}_{\sigma}$,
so  $\eta_{PP}$ must be nonzero.
For the Poincar\'{e} group the canonical dimensions of 
the momentum and 
the Lorentz generator are 1 and  0 respectively.
A nondegenerate partner of the Lorentz generator has
the canonical dimension 2,
so that the sum of canonical dimensions of a nondegenerate pair
is 2.  
For the manifest covariance including the curvature tensors 
the Lorentz generator must be involved.

At first we present the general method to make a non-abelian  group  
to be nondegenerate for a symmetric space given by a coset group G/H. 
\begin{enumerate}
   \item {For a coset group G/H$_0$ a subgroup
   H$_0$ corresponds to the Lorentz group  
   generated by  $h_0$ and G/H$_0$ is generated by $k$. 
They satisfy the following algebra,
 \bea
&\lbrack h_0,h_0]=h_0~,~[h_0,k]=k~,~[k,k]=h_0~~~.& \label{hok}
\eea}
   \item {Introduce another copy of the subgroup  H$_1$ \cite{Bonanos:2009wy}
in order to make G$\times$H$_1$ to be nondegenerate.
   H$_1$ is generated by $h_1$,
   \bea
&\lbrack h_1,h_1]=h_1~~~~.& \label{h1h1}
\eea    }
   \item{Make nondegenerate pair $h$ and $\check{h}$ 
     by linear combinations of $h_0$ and $h_1$ as
   \bea
&{\renewcommand{\arraystretch}{1.6}\left\{\begin{array}{l}
h_0+ h_1=h\\
h_0-h_1=\check{h}\\
k\to k/\sqrt{2}
\end{array}\right.}
~\Rightarrow~
{\renewcommand{\arraystretch}{1.6}\left\{\begin{array}{l}
\lbrack h,h]=h~,~[h,\check{h}]=\check{h}~,~\lbrack \check{h},\check{h}]=h\\
\lbrack h,k]=k~,~[k,k]=h+\check{h}~,~[\check{h} ,k]=k
\end{array}\right.}~~.&
\eea
$h$ and $\check{h}$ are generators 
  of H and $\check{\rm H}$ which are subgroups of G$\times$H$_1$.
}
\item{Non-zero components of the 
 nondegenerate group  metric are
\bea
{\rm tr}(kk)=
\eta_{kk}~,~
{\rm tr}(h\check{h})=
\eta_{h\check{h}}~~.
\eea
The structure constant lowered by the nondegenerate group  metric becomes totally antisymmetric
\bea
f_{hh\check{h}}=f_{\check{h}\check{h}\check{h}}=
f_{hkk}=f_{\check{h}kk}={\bf 1}~~~.
\eea
}
\end{enumerate}
\par
\vskip 6mm

\subsubsection{Doubled Lie algebra}

The gravitational field is described by a closed string 
which has the left and right moving modes.
We  begin by the sigma model Lagrangian for a closed string
\bea
{\cal L}&=&-\frac{1}{2}\left(\sqrt{-h}h^{ij}\partial_i x^m\partial_j x^n G_{mn}
+\epsilon^{ij}\partial_i x^m\partial_j x^n B_{mn}\right)~~~.
\eea
In the conformal gauge, the Lagrangian is rewritten in
 the doubled basis
 $\partial_\pm x^{ m}=\frac{1}{\sqrt{2}}(\partial_\tau\pm \partial_\sigma)x^{ m}$ with the two vielbein fields  $e_{\rm a}{}^{m}$ and $e'_{m{\rm a}}$ 
 as \cite{Siegel:1993xq}
 \bea
{\cal L}_{\rm conformal~gauge}&=&
\frac{1}{2}j^{\underline{\rm a}}~{\eta}_{\underline{\rm ab}}~j^{\underline{\rm b}}\nn
~~,~~
G_{ mn}+B_{ mn}=e_{ m}{}^{\rm a}e'{}_{n{\rm a}}
\\
j^{\underline{\rm a}}&=&
{\renewcommand
{\arraystretch}{1.2}
\left\{\begin{array}{cl}
j^{\rm a}&=\partial_+ x^{ m}e_{ m}{}^{\rm a}\\
j_{\rm a}&=\partial_- x^{ m}e'{}_{ m}{}_{\rm a}
\end{array}\right.}\label{pmc}~,~~
{\eta}_{\underline{\rm ab}}=
\left(\begin{array}{cc}0&\delta_{\rm a}^{\rm b}
\\\delta_{\rm b}^{\rm a}&0
\end{array}
\right)~~~.
\eea
The left and right currents are written in term of the canonical momentum
 $ p_{ m}\equiv \frac{\partial {\cal L}}{\partial 
 \partial_\tau x^{ m}}=G_{mn}\partial_\tau x^n
 +B_{nm}\partial_\sigma x^n$ 
\bea
j^{\underline{\rm a}}&=&{\renewcommand
{\arraystretch}{1.2}
\left\{\begin{array}{cl}
j^{\rm a}&=
\frac{1}{\sqrt{2}}\left(\eta^{\rm ab}e_{\rm b}{}^{ m}
(p_{ m}+B_{ mn}\partial_\sigma x^{ n})
+\partial_\sigma x^{ n}e_{ n}{}^{\rm a}\right)\\
j_{\rm a}&=\frac{1}{\sqrt{2}}\left(
e'{}_{ l{\rm a}}G^{ lm}(p_{m}+B_{ mn}\partial_\sigma x^{n})
-\partial_\sigma x^{ n}e'{}_{n}{}_{\rm a}\right)
\end{array}\right.}~~~\label{delrma}
\eea
with $G^{mn}=e_{\rm a}{}^{m}\eta^{\rm ab}e_{\rm b}{}^{n}$
and $e_{\rm a}{}^me_m{}^{\rm b}=\delta_{\rm a}^{\rm b}$.
The basis of the doubled space are essentially the left and right moving modes.

On the other hand the Hamiltonian with the 
 two dimensional diffeomorhism invariance
 is given by
\bea
{\cal H}&=&
\frac{1}{\sqrt{-h}h^{00}}{\cal H}_{\tau}-\frac{h^{01}}{h^{00}}{\cal H}_\sigma~,~
\left\{{\renewcommand{\arraystretch}{1.6}\begin{array}{l}
{\cal H}_\sigma=
\frac{1}{2}\dd_{\underline{\rm a}}~{\eta}^{\underline{\rm ab}}~\dd_{\underline{\rm b}}=\frac{1}{2}\dron{\scriptscriptstyle\bo}\dd_{\underline{\rm m}}~{\eta}^{\underline{\rm mn}}~\dron{\scriptscriptstyle\bo}\dd_{\underline{\rm n}}
\\
{\cal H}_\tau=\frac{1}{2}\dd_{\underline{\rm a}}~\hat{\eta}^{\underline{\rm ab}}~\dd_{\underline{\rm b}}=\frac{1}{2}\dron{\scriptscriptstyle\bo}\dd_{\underline{\rm m}}~{\cal M}^{\underline{\rm mn}}~\dron{\scriptscriptstyle\bo}\dd_{\underline{\rm n}}
\end{array}}\right.\nn
\\
\dd_{\underline{\rm a}}&=&
e_{\underline{\rm a}}{}^{\underline{\rm m}}\dron{\scriptscriptstyle\bo}\dd_{\underline{\rm m}}~,~
\dron{\scriptscriptstyle\bo}\dd_{\underline{\rm m}}=
\left(\begin{array}{c}p_{ m}\\\partial_\sigma x^{ m}\end{array}\right)
~,~ e_{\underline{\rm a}}{}^{\underline{\rm m}}{\eta}^{\underline{\rm ab}}e_{\underline{\rm b}}{}^{\underline{\rm n}}
 ={\eta}^{\underline{\rm mn}}~,~
 e_{\underline{\rm a}}{}^{\underline{\rm m}}{\eta}_{\underline{\rm mn}}e_{\underline{\rm b}}{}^{\underline{\rm n}}
 ={\eta}_{\underline{\rm ab}}
 ~\label{orthogonal}\\
\hat{\eta}^{\underline{\rm  ab}}&=&
\left(\begin{array}{cc}\eta^{\rm ab}&
\\&\eta_{\rm ab}\end{array}\right)~,~{\cal M}^{\underline{\rm mn}}=
e_{\underline{\rm a}}{}^{\underline{\rm m}}\hat{\eta}^{\underline{\rm ab}}e_{\underline{\rm b}}{}^{\underline{\rm n}}
={\renewcommand{\arraystretch}{1.2}
\left(\begin{array}{cc}G^{mn}&G^{ ml}
B_{ln}\\
-B_{ ml}G^{ ln}&
G_{ mn}-B_{ ml}G^{ lk}B_{kn}
\end{array}
\right)}\nn
\eea
The conformal gauge is given by $\frac{1}{\sqrt{-h}h^{00}}=1$ and $\frac{h^{01}}{h^{00}}=0$.
The covariant derivatives in arbitrary curved backgrounds
 are written as the vielbein multiplied on the flat space covariant derivative as 
$\dd_{\underline{\rm a}}=e_{\underline{\rm a}}{}^{\underline{\rm m}}\dron{\scriptscriptstyle\bo}\dd_{\underline{\rm m}}$.
The doubled vielbein field $e_{\underline{\rm a}}{}^{\underline{\rm m}}$  satisfies the orthogonal condition
\bref{orthogonal}, so it is an element of the coset 
\bea
\frac{\rm O(d,d)}{{\rm SO}({\rm d}-1,1)\times{\rm SO}(1,{\rm d}-1)}~~~.
\label{cosetdd}
\eea
The number of physical degres of freedom for  $G_{mn}$ and $B_{mn}$ is d$^2$ which is  the number of the dimensions of the coset in \bref{cosetdd}.
While  $G_{mn}+B_{mn}$ is transformed fractional linearly, 
the vielbein $e_{\underline{\rm a}}{}^{\underline{\rm m}}$ is transformed linearly under the O(d,d) T-duality symmetry transformation, 
$e_{\underline{\rm a}}{}^{\underline{\rm m}}
\to h_{\underline{\rm a}}{}^{\underline{\rm b}}e_{\underline{\rm b}}
{}^{\underline{\rm n}}\Lambda_{\underline{\rm m}}{}^{\underline{\rm n}}$
with  $\Lambda^T{\eta}\Lambda={\eta}$ and 
$h^T\hat{\eta}h=\hat{\eta}$.  

An O(d,d) T-duality transformation which 
produces nonzero  $B$ field, ~$B_{mn}=0\to B_{mn}=\lambda_{mn}~$, is given by
\bea
&&e_{\underline{\rm a}}{}^{\underline{\rm m}}
=\left(\begin{array}{cc}e^{-1}&\\&e^T\end{array}\right)
~,~
\Lambda=\left(\begin{array}{cc}1&-\lambda\\&1
\end{array}\right)~,~\lambda^T=-\lambda~\nn\\
&&\to e_{\underline{\rm a}}
{}^{\underline{\rm n}}(\Lambda^{-1})_{\underline{\rm n}}{}^{\underline{\rm m}}
=\left(\begin{array}{cc} e^{-1}&\\& e^{T}
\end{array}\right)
\left(\begin{array}{cc}1&\lambda\\ &1\end{array}\right)~~
\eea
where indices are omitted as $e,\lambda$ for $e_{m}{}^a,\lambda_{mn}$
for simpler notation.
Another O(d,d)$\ni \Lambda$ transformation which interchanges the momenta and the winding modes the vielbein is transformed as:
\bea
&&\dron{\scriptscriptstyle\bo}\dd_{\underline{\rm m}}\to \Lambda_{\underline{\rm m}}{}^{\underline{\rm n}}\dron{\scriptscriptstyle\bo}\dd_{\underline{\rm n}}~,~e_{\underline{\rm a}}{}^{\underline{\rm m}}
\to h_{\underline{\rm a}}{}^{\underline{\rm b}}e_{\underline{\rm b}}
{}^{\underline{\rm n}}(\Lambda^{-1})_{\underline{\rm m}}{}^{\underline{\rm n}}~\\
&&e_{\underline{\rm a}}{}^{\underline{\rm m}}
=\left(\begin{array}{cc}e^{-1}&\\&e^T\end{array}\right)
\left(\begin{array}{cc}1&B\\&1\end{array}\right)
~,~
\Lambda=\left(\begin{array}{cc}&(\lambda^{-1})^T\\\lambda&
\end{array}\right)~,~
h=\left(\begin{array}{cc}&1\\1&
\end{array}\right)\nn\\
&&\to~h_{\underline{\rm a}}{}^{\underline{\rm b}}e_{\underline{\rm b}}
{}^{\underline{\rm n}}(\Lambda^{-1})_{\underline{\rm n}}{}^{\underline{\rm m}}
=\left(\begin{array}{cc}(\lambda e)^T&\\&(\lambda e)^{-1}
\end{array}\right)
\left(\begin{array}{cc}1&\\\lambda B\lambda^T&1\end{array}\right)\nn
\eea
where indices are omitted as $B,G$ for $B_{mn},G_{mn}$.
This simple transformation rule corresponds to the following
transformation rules of $G_{mn}$ and $B_{mn}$ as
\bea
&&
{\renewcommand{\arraystretch}{1.6}
\left\{\begin{array}{ccl}
G_{mn}&\to&
(\lambda^{-1})^T(G-BG^{-1}B)^{-1}\lambda^{-1}\\
B_{mn}&\to&
\left[(\lambda^{-1})^TG^{-1}B(G-BG^{-1}B)^{-1}\lambda^{-1}\right]_{[mn]/2}\\
\end{array}\right.}
\eea
with antisymmetrizing indices, ${\cal O}_{[mn]/2}=({\cal O}_{mn}-{\cal O}_{nm})/2 $.
They are generalizations of the Buscher's transformation rule.

It is known that doubled coordinates manifest T-duality symmetry,
and the physical degrees of freedom is a half of it.
The section condition is usually considered as
$\partial_{\rm m}\partial^{\rm m}=0$
where $\partial_{\rm m}=\frac{\partial}{\partial x^{\rm m}}$ and 
$\partial^{\rm m}
=\frac{\partial}{\partial y_{\rm m}}$,
and it is imposed on the spacetime field weakly as
$\partial_{\rm m}\partial^{\rm m}\Psi(x^{\rm m},y_{\rm m})=0$
and strongly 
$\partial_{\rm m}\Phi(x^{\rm m},y_{\rm m})\partial^{\rm m}\Psi(x^{\rm m},y_{\rm m})=0$.
The $y_{\rm m}$-independence satisfies the section condition
and the theory reduces to the usual coordinate space theory. 
This condition is the $\sigma$-diffeomorphism invariance constraint
${\cal H}_\sigma=\partial_{\rm m}\partial^{\rm m}=0$
for a string on the worldsheet.
The $\sigma$-diffeomorphism invariance constraint 
is imposed on fields as a matrix element of the second quantized level,
$\langle \Phi|{\cal H}_\sigma|\Psi\rangle=0$.
In other words fields in the target space governed by the string theory
should  be  $\sigma$-diffeomorphism invariant. 

The doubled momenta  $\dron{\scriptscriptstyle\bo}\dd_{\underline{m}}=P_{\underline{m}}=(P_{ m},P_{m'})$ are independent, so we have doubled coordinates.
Then we impose dimensional reduction constraints to reduce a half.
We do not impose gauge fixing conditions on  
spacetime fields $\frac{\partial}{\partial y_{\rm m}}\Psi=0$,
and they are written as $P_{ m}=(p_m+\partial_\sigma x^m)/\sqrt{2}$ and 
$P_{m'}=(p_{m'}-\partial_\sigma x^{m'})/\sqrt{2}$ in a flat space
with $x^m=(x^{\rm m}+y_{\rm m})/\sqrt{2}$ and $x^{m'}=(x^{\rm m}-y_{\rm m})/\sqrt{2}$.
 The dimensional reduction constraints are 
 first class, so  the local gauge symmetry 
and all doubled coordinates are preserved.
Therefore the T-duality covariant general coordinate invariance of the stringy  gravity is manifest.

The dimensional reduction constraints are made from the right-invariant one form,
while the local geometry is made from the left-invariant
one form so that the auxiliary coordinates are reduced by the dimensional reduction constraints
without modifying the local geometry.
In order to construct the left-invariant one form and the right-invariant one form for both left and right moving modes we double the whole group
\bea
 {\rm G}\to {\rm G}\times{\rm  G}'~~~.
 \eea
A group element of the direct product of these groups G$\times$G'$\ni \underline{g}=g(Z^M)g(z^{M'})$ gives both the left and right moving modes of the left-invariant and the right-invariant currents; $\underline{g}^{-1}d\underline{g}=
g^{-1}dg(Z)+g'^{-1}dg'(Z')=iJ(Z)+iJ(Z')$ and 
$d\underline{g}\underline{g}^{-1}=
dgg^{-1}(Z)+dg'g'^{-1}(Z')=i\tilde{J}(Z)+i\tilde{J}(Z')$.
For the RR background this factorization is nontrivial as seen later.

\par\vskip 6mm

\subsection{Affine Lie algebras }

Let us go back to the procedure of the manifestly T-dual formulation
in arbitrary group manifolds.
We begin by a Lie algebra generated by $G_I$  
 \bea
\lbrack G_{I},G_{J}]=
i{f}{}_{IJ}{}^{K}G_{K}\label{GGfGflat}~,~
~{\rm tr}(G_IG_J)=\eta_{IJ}~,~\det \eta_{IJ}\neq 0~~.
\eea
For the Lie algebra in \bref{GGfGflat} 
its group element $g$  is parametrized by $Z^M$
where the number of Lie algebra generators  
$G_I$ is equal to the number of the parameters $Z^M$.
We extend it to affine  Lie algebras as string algebras.
The coordinates $Z^M$'s  
are functions of  the two-dimensional worldsheet coordinates.
Generators of affine Lie algebras are constructed from
the left and right-invariant currents and the particle
covariant derivative and the particle symmetry generator.
\begin{itemize}
  \item{
\begin{description}
  \item[Left-invariant one form and the particle covariant derivative] 
\end{description}
The left-invariant one form $J$
satisfies the Maurer-Cartan equation, and
the covariant derivative $\nabla_I$ 
satisfies the following equation 
\bea
&g^{-1}dg=iJ^{}{}^IG_I~~,~~J^{}{}^I=dZ^MR^{}{}_{M}{}^I~~
\Rightarrow~~dJ^{}{}^I=-\frac{1}{2}f^{}{}_{JK}{}^IJ^{}{}^J\wedge J^{}{}^K~~~&\nn\\
&\nabla^{}{}_I=(R^{-1})_I{}^M\frac{1}{i}\partial_M~\Rightarrow~~
(R^{-1})_{[I}{}^M\nabla_{J]}R_{M}{}^{K}=if_{IJ}{}^K~~.&\label{dJJJ}
\eea
} 
\item{
\begin{description}
  \item[Right-invariant current and the particle symmetry generator] 
\end{description}
The right-invariant one form $\tilde{J}$
satisfies the Maurer-Cartan equation, 
and the symmetry generator $\tilde{\nabla}_I$ satisfies the following equation
\bea
&dgg^{-1}=i\tilde{J}^{}{}^IG_I~~,~~\tilde{J}^{}{}^I=dZ^ML^{}{}_{M}{}^I~~
\Rightarrow~~d\tilde{J}^{}{}^I=\frac{1}{2}f^{}{}_{JK}{}^I\tilde{J}^{}{}^J\wedge \tilde{J}^{}{}^K~~~&\nn\\
&\tilde{\nabla}^{}{}_I=(L^{-1})_I{}^M\frac{1}{i}\partial_M~~\Rightarrow~~
(L^{-1})_{[I}{}^M\nabla_{J]}L_{M}{}^{K}=-if_{IJ}{}^K~~.&\label{dJJJ}
\eea
}  

  \item{   
\begin{description}
  \item[Algebras by particle covariant derivative and symmetry generator] 
\end{description}
The covariant derivative and the symmetry generator together with
$J^{}_1{}^{I}=\partial_\sigma Z^M R_M{}^I$  and
$\tilde{J}^{}_1{}^{I}=\partial_\sigma Z^M L_M{}^I$ 
satisfy the following affine Lie algebras:
 \bea
&&{\renewcommand{\arraystretch}{1.6}
\left\{\begin{array}{ccl}
\lbrack\nabla^{}{}_{I}(1),\nabla^{}{}_{J}(2)]&=&-i
{f}^{}{}_{IJ}{}^{K}\nabla^{}{}_{K}\delta(2-1)\\
\lbrack\nabla^{}{}_{I}(1),J^{}_1{}^{J}(2)]&=&-iJ^{}_1{}^{K}
{f}^{}{}_{KI}{}^{J}\delta(2-1)-i\delta_I^J\partial_\sigma\delta(2-1)
\\
\lbrack J^{}_1{}^{I}(1),J^{}_1{}^{J}(2)]&=&0
\end{array}\right.}\label{nabla}~~~.
\\
&&{\renewcommand{\arraystretch}{1.6}
\left\{\begin{array}{ccl}
\lbrack\tilde{\nabla}^{}{}_{I}(1),\tilde{\nabla}^{}{}_{J}(2)]&=&i
{f}^{}{}_{IJ}{}^{K}\tilde{\nabla}^{}{}_{K}\delta(2-1)\\
\lbrack\tilde{\nabla}^{}{}_{I}(1),\tilde{J}^{}_1{}^{J}(2)]&=&i\tilde{J}^{}_1{}^{K}
{f}^{}{}_{KI}{}^{J}\delta(2-1)+i\delta_I^J\partial_\sigma\delta(2-1)
\\
\lbrack \tilde{J}^{}_1{}^{I}(1),\tilde{J}^{}_1{}^{J}(2)]&=&0
\end{array}\right.}
\label{sym}~~~.\\
&&{\renewcommand{\arraystretch}{1.6}
\left\{\begin{array}{ccl}
\lbrack{\nabla}^{}{}_{I}(1),\tilde{\nabla}^{}{}_{J}(2)]&=&0\\
\lbrack\tilde{\nabla}^{}{}_{I}(1),{J}^{}_1{}^{J}(2)]&=&-i
M_I{}^J(2)\partial_\sigma\delta(2-1)
\\
\lbrack \nabla_{I}(1),\tilde{J}^{}_1{}^{J}(2)]&=&-i
(M^{-1})_I{}^J(2)\partial_\sigma\delta(2-1)
\end{array}\right.}
\label{nablasym}~~~
\eea
with
\bea
&&M_I{}^J=(L^{-1})_I{}^MR_M{}^J~~,~~\tilde{J}^IM_I{}^K=J^K~~,~~\tilde{\nabla}_I=M_I{}^K\nabla_K~~\nn\\
&&\eta_{IJ}=M_I{}^LM_J{}^K\eta_{LK}~~,~~
f_{IJK}=M_I{}^LM_J{}^PM_K{}^Qf_{LPQ}~~
.\label{MMeta}
\eea 
$\sigma_1$ and $\sigma_2$ are abbreviated as $1$ and $2$,
and  $\delta(2-1)=\delta(\sigma_2-\sigma_1)$ and
$\partial_\sigma\delta(2-1)=\partial_{\sigma_2}\delta(\sigma_2-\sigma_1)$.  
From the relation between the left-invariant one form and the
right-invariant one form $g^{-1}\tilde{J}g=J$ $\to$
$L_M{}^Ig^{-1}G_Ig=R_M{}^IG_I$,
the nondegenerate group metric 
$\eta_{IJ}={\rm tr}(G_IG_J)={\rm tr}(g^{-1}G_Igg^{-1}G_Jg)$ 
leads to that  the matrix $M_I{}^J$ satisfies the orthogonal condition and invariance of the structure constant \bref{MMeta}. 
}
\end{itemize}

The string covariant derivative  $\dd^{}{}_I(\sigma)$ is 
constructed with the $B$ field from the particle
 covariant derivative  $\nabla_I(\sigma)$ and the $\sigma$ component of
 the  left-invariant current $J_1{}^I(\sigma)$.
 The string symmetry generator $\tilde{\dd}^{}{}_I(\sigma)$ is 
constructed with the $\tilde{B}$ field from the particle
 symmetry generator $\tilde{\nabla}_I(\sigma)$
  and the $\sigma$ component of
the right-invariant current $\tilde{J}_1{}^I(\sigma)$ as;

\begin{itemize}
  \item {
  \begin{description}
  \item[Covariant~derivative] \footnote{The coefficient $\frac{1}{2}$ arises from the normalization of the Schwinger term in the affine Lie algebra.
The same normalization of the Schwinger term is satisfied by 
$\frac{1}{\sqrt{2}}\left(\nabla^{}{}_I+J_1^{}{}^K(\eta_{KI}+B^{}{}_{KI})\right)
$.}
\end{description}
\bea
&\dd^{}{}_I=\nabla^{}{}_I+\frac{1}{2}J_1^{}{}^K(\eta_{KI}+B^{}{}_{KI})~~~\label{ddDJB}
&
\eea
}
\item {
  \begin{description}
  \item[Symmetry~generator] 
\end{description}
\bea
&\tilde{\dd}^{}{}_I=\tilde{\nabla}^{}{}_I
+\frac{1}{2}\tilde{J}_1^{}{}^K(-\eta_{KI}+\tilde{B}^{}{}_{KI})&
\label{symgenflat4}
\eea}
\end{itemize}
The affine extension of the Lie algebra 
\bref{GGfGflat} is performed  using \bref{nabla}, \bref{sym} and \bref{nablasym}.
\begin{itemize}
\item {
  \begin{description}
  \item[Affine Lie algebras ] 
\end{description}
\bea
\lbrack {\dd}_{I}(1),{\dd}_{J}(2)]&=&-if_{IJ}{}^{K}{\dd}_{K}\delta(2-1)
-i\eta_{IJ}\partial_\sigma\delta(2-1)\nn\\
\lbrack  {\tilde{\dd}}_{I}(1), {\tilde{\dd}}_{J}(2)]&=&if_{IJ}{}^{K} {\tilde{\dd}}_{K}\delta(2-1)+i\eta_{IJ}\partial_\sigma\delta(2-1)
\label{stringy}
\\
\lbrack {\dd}_{I}(1),{\tilde{\dd}}_{J}(2)]&=&0~\nn
\eea
}
\end{itemize}

The antisymmetric tensor $B_{IJ}$ field in the covariant derivative
must satisfy the following equation \cite{Hatsuda:2015cia}
\bea
i{\nabla}_{[I}{B}_{JK]}-{f}_{[IJ|}{}^L{B}_{L|K]} =
2{f}_{IJK}\label{dBH}
~~~.
\eea
Another antisymmetric tensor $\tilde{B}_{IJ}$ field in the symmetry generator
is related to $B_{IJ}$ from \bref{MMeta} as
\bea
\tilde{B}_{IK}=M_I{}^JB_{JL}M_K{}^L~~.\label{BtilB}
\eea
The two form $B$ gives the Wess-Zumino term for a fundamental string
\bea
&&B=\frac{1}{2}dZ^M\wedge dZ^N B_{MN}=\frac{1}{2}J^I \wedge J^J B_{IJ}
=\frac{1}{2}\tilde{J}^I \wedge \tilde{J}^J \tilde{B}_{IJ}
\nn\\
&&B_{MN}=R_M{}^IR_N{}^JB_{IJ}=L_M{}^IL_N{}^J\tilde{B}_{IJ}~~~.
\eea
The three form $H=dB$ is calculated with \bref{dBH} as
\bea
&&H=dB=\frac{1}{3!}dZ^M\wedge dZ^N \wedge dZ^L H_{MNL}
=\frac{1}{3!}J^I \wedge J^J \wedge J^K f_{IJK}
=\frac{1}{3!}\tilde{J}^I \wedge \tilde{J}^J \wedge \tilde{J}^K f_{IJK}
\nn\\
&&H_{MNP}=R_M{}^IR_N{}^JR_P{}^Kf_{IJK}=L_M{}^IL_N{}^JL_P{}^Kf_{IJK}~~~.\label{HH}
\eea
It is also denoted that the condition on $B_{IJ}$ in \bref{dBH} 
is expressed as $dB=H$ where $H$ is given in \cite{Hatsuda:2015cia}. 
$B$ is determined from it up to its  gauge freedom $d\lambda$. 
The existence of the solution is guaranteed by 
$dH = 0$, which is proven using Maurer-Cartan equations.

The  $\sigma$ diffeomorphsim generator
is defined by bilinears of  the covariant derivatives contracted with the nondegenerate group metric as
\bea
{\cal H}_\sigma=\frac{1}{2}\dd_I^{}{\eta}^{IJ}\dd_J^{}~~~.
\eea
For a field $\Phi$ which is a function of 
the group manifold coordinates,
the $\sigma$ derivative of $\Phi$ is determined as
\bea
\partial_\sigma\Phi&=&i\int d\sigma'\lbrack{\cal H}_\sigma(\sigma'),\Phi]=
\dd_I^{}\eta^{IJ}(i\nabla_J^{} \Phi)~~~
.\label{sigmader}
\eea
If a field $\Phi$ is a function of 
both phase space coordinates $(Z^M,~\frac{1}{i}\partial_M)$, then the derivative  $(i\nabla_J^{} \Phi)$ is replaced by
the commutator as $[i\dd_J, \Phi]$ in \bref{sigmader}.

Let us  consider  a space defined  
by the affine Lie algebra
generated by the covariant derivative in the first line of \bref{stringy}.
Two vectors in the space, 
$\hat{\Lambda}_i=\Lambda_i{}^I(Z^M) \dd_I(\sigma)$ with $i=1,2$, 
satisfy the commutator  as
\bea
&&\lbrack {\Lambda}_1{}^I\dd_I(1),{\Lambda}_2{}^J\dd_J(2)]\nn\\
&&~~~=-i\Lambda_{12}{}^I\dd_{I}\delta(2-1)-i\{
(\frac{1}{2}+{\cal K})\Psi_{(12)}(1)+(\frac{1}{2}-{\cal K})\Psi_{(12)}(2)
\}\partial_\sigma\delta(2-1)\nn\\
&&~~~\Lambda_{12}{}^I=\Lambda_{[1|}{}^K(i\nabla_K\Lambda_{|2]}{}^I)
-\frac{1}{2}\Lambda_{[1|}{}^K(i\nabla^I\Lambda_{|2]}{}_K)
+\Lambda_1{}^J\Lambda_2{}^Kf_{JK}{}^I
-{\cal K}(i\nabla^I\Psi_{(12)})\nn\\
&&~~~\Psi_{(12)}=\Lambda_1{}^I\Lambda_{2}{}^J\eta_{IJ}~~~\label{Lam12}
\eea
where $\sigma$ derivative is calculated by \bref{sigmader}.
The factors ``$i$"'s come from the definition of covariant derivative
$\nabla_I=R_I{}^M\frac{1}{i}\partial_M$. 
There is an ambiguity with parameter ${\cal K}$ caused from the 
Schwinger term including $\partial_\sigma\delta(2-1)$. 
The regular part of the algebra is a generalization of the 
 ``C-bracket" 
\bea
\left(\lbrack{\Lambda}_1,{\Lambda}_2]_{\rm T}\right)^I=
-i{\Lambda}_{12}{}^I~~~~~~~\label{Tbracket}
\eea
where we put ``T" which stands for T-duality.
The expression of  $\Lambda_{12}$ depends on the value of ${\cal K}$ as
\bea
\Lambda_{12}{}^I=\left\{
{\renewcommand{\arraystretch}{1.6}
\begin{array}{lc}
\Lambda_{[1|}{}^K(i\nabla_K\Lambda_{|2]}{}^I)
-\frac{1}{2}\Lambda_{[1|}{}^K(i\nabla^I\Lambda_{|2]}{}_K)
+\Lambda_1{}^J\Lambda_2{}^Kf_{JK}{}^I&\cdots {\cal K}=0\\
\Lambda_{1}{}^K(i\nabla_K\Lambda_{2}{}^I)
+\Lambda_{2}{}^K(i\nabla_{[J|}\Lambda_{1}{}_{|K]})\eta^{JI}
+\Lambda_1{}^J\Lambda_2{}^Kf_{JK}{}^I&\cdots {\cal K}=-\frac{1}{2}
\end{array}}\right.~~~.
\eea 
For the case with ${\cal K}=0$ it is antisymmetric under
  $1\leftrightarrow 2$ interchanging, while the case with ${\cal K}=-1/2$ gives the usual gauge transformation rules.
The Jacobi identity of the T-bracket is not satisfied in general
because of  luck of the contribution from the Schwinger term.
The Jacobi identity of the affine algebra \bref{Lam12} is the Bianchi identity 
giving a condition on $\Lambda_{12}$.

\par
\vskip 6mm
\section{Flat space}

\subsection{Dilatation  operator and $B$ field }
We begin by the Poincar\'{e} algebra
as a flat space, then introduce the nondegenerate partner
of the Lorentz generator following to the previous section.
The nondegenerate Poincar\'{e} algebra  is
generated by $G_M$.
In this case there exists a dilatation operator
 $\hat{N}$  and 
 the  canonical  dimensions of  generator $G_M$ is $n_M$ as 
\bea
\lbrack G_{M},G_{N}]=
i\don{\scriptscriptstyle\bo}{f}{}_{MN}{}^{L}G_{L}~~,~~
 \lbrack \hat{N},G_M]=iN_M{}^NG_N=in_MG_M~~~.
\eea
The generator of the nondegenerate Poincar\'{e} algebra, 
$G_M=(s_{mn},p_m,\sigma^{mn})$, 
and the dilatation operator $\hat{N}$ satisfy the following algebra
\bea
 {\renewcommand{\arraystretch}{1.6}
\begin{array}{c}
\lbrack s_{mn},s_{lk}]=i\eta_{[k|[m}s_{n]|l]}~,~
\lbrack s_{mn},p_{l}]=ip_{[m}\eta_{n]l}\\
\lbrack s_{mn},\sigma_{lk}]=i\eta_{[k|[m}\sigma_{n]|l]}~,~
\lbrack p_{m},p_{n}]=i\sigma_{mn}\\
\lbrack  \hat{N},s_{mn}]=0~,~
\lbrack \hat{N},p_{m}]=ip_{m}~~,~~
\lbrack \hat{N}, \sigma^{mn}]=2i\sigma^{mn}
\end{array}}\label{Poinflat}~~~.
\eea
The nondegenerate group metric is
\bea
&&~~~~~~~~s~~~p~~~\sigma\nn\\
\eta_{IJ}&=&\begin{array}{c}s\\p\\\sigma\end{array}
\left(\begin{array}{ccc}&&1\\&1&\\1&&\end{array}
\right)
\eea 
The sum of the canonical dimensions of the
nondegenerate pair is 2; $(n_I+n_J)\eta_{IJ}=2\eta_{IJ}$.
The Jacobi identity among $\hat{N}$ and two  $G_M$'s leads to an identity
\bea
\don{\scriptscriptstyle\bo}{f}_{MN}{}^KN_K{}^L+N_{[M}{}^K\don{\scriptscriptstyle\bo}{f}_{N]K}{}^L=(n_L-n_M-n_N)\don{\scriptscriptstyle\bo}{f}_{MN}{}^L=0~~~,
\label{NGG}
\eea
so the sum of the canonical dimensions of the lowered indices of the non-zero component of the structure constant
is also 2; $(n_M+n_N+n_L)\don{\scriptscriptstyle\bo}{f}_{MNL}=2\don{\scriptscriptstyle\bo}{f}_{MNL}$.
This identity gives a constant $B$  field  solution 
 of the equation \bref{dBH}
for the nondegenerate Poincar\'{e} group as
\bea
\don{\scriptscriptstyle\bo}B{}_{NM}=-\frac{1}{2}N_{[N|}{}^L\eta_{L|M]}
=\frac{1}{2}(-n_N+n_M)\eta_{NM}~~~.\label{Bflatconstant}
\eea
As a result the stringy covariant derivative for the flat space 
$\dron{\scriptscriptstyle\bo}\dd{}_{{N}}$
 is written in terms of the particle covariant derivative
 $\dron{\scriptscriptstyle\bo}\nabla{}_{{N}}$
 and the $\sigma$-component of the left-invariant current
 $\dron{\scriptscriptstyle\bo}J_1{}^{{N}}$ in the flat space
 as\footnote{The coefficient $\frac{1}{2}$ arises from the normalization of the Schwinger term in the affine Lie algebra \bref{flatcovIJ}. 
Another normalization gives the usual stringy covariant derivative
\bea
\dron{\scriptscriptstyle\bo}\dd{}_{M}&=&
\frac{1}{\sqrt{2}}(\dron{\scriptscriptstyle\bo}{\nabla}{}_{M}
+{n_M} \don{\scriptscriptstyle\bo}{J}_{M})~~~.
\eea
}
\bea
\dron{\scriptscriptstyle\bo}\dd{}_{M}&=&\don{\scriptscriptstyle\bo}{\nabla}{}_{M}
+\displaystyle\frac{1}{2}\don{\scriptscriptstyle\bo}{J}_1{}^{L}
(\eta_{LM}+\don{\scriptscriptstyle\bo}B{}_{LM})~=~
\don{\scriptscriptstyle\bo}{\nabla}{}_{M}
+\displaystyle\frac{n_M}{2}  \don{\scriptscriptstyle\bo}{J}_{1;M}\label{Jbo}
\eea
with $\don{\scriptscriptstyle\bo}J_{1;M}\equiv \don{\scriptscriptstyle\bo}J_1{}^L\eta_{LM}$
. 
It satisfies the affine nondegenerate Poincar\'{e} 
 algebra 
  \bea
\lbrack \dron{\scriptscriptstyle\bo}\dd{}_{M}(1),
\dron{\scriptscriptstyle\bo}\dd{}_{N}(2)]
=-i\don{\scriptscriptstyle\bo}{f}{}_{MN}{}^{L}
\dron{\scriptscriptstyle\bo}\dd{}_{L}\delta(2-1)
-i\eta_{MN}\partial_\sigma\delta(2-1)~~~.\label{flatcovIJ}
\eea

\par
\vskip 6mm
\subsection{Affine Poincar\'{e} algebras}

Next the nondegenerate Poincar\'{e} algebra is doubled
as described in the previous section
\bea
G_M&\to& G_{\underline{M}}=(G_M,~G_{M'}) \nn\\
\don{\scriptscriptstyle\bo}f_{MN}{}^L&\to& \don{\scriptscriptstyle\bo}f_{\underline{MN}}{}^{\underline{L}}=(\don{\scriptscriptstyle\bo}f_{MN}{}^L, ~
\don{\scriptscriptstyle\bo}f_{M'N'}{}^{L'}=-\don{\scriptscriptstyle\bo}f_{MN}{}^L)~\label{GGp}\\
\eta_{MN}&\to&{\renewcommand{\arraystretch}{1.6}
\left\{\begin{array}{l}
\eta_{\underline{MN}}=(\eta_{MN},~\eta_{M'N'}=-\eta_{MN})\\
\hat{\eta}_{\underline{MN}}=(\eta_{MN},~\hat{\eta}_{M'N'}=\eta_{MN})
\end{array}\right.}
~~~\nn .
\eea
Covariant derivatives and symmetry generators
for the nondegenerate doubled Poincar\'{e} algebra
are given as follows.
\begin{itemize}
  \item{\begin{description}
  \item[Flat covaiant derivatives : 
] $\dron{\scriptscriptstyle\bo}\dd{}_{\underline{M}}=\don{\scriptscriptstyle\bo}{\nabla}{}_{\underline{M}}
+\displaystyle\frac{1}{2}\don{\scriptscriptstyle\bo}{J}_1{}^{\underline{L}}
(\eta_{\underline{LM}}+\don{\scriptscriptstyle\bo}B{}_{\underline{LM}})=
(\dron{\scriptscriptstyle\bo}\dd{}_{{M}},\dron{\scriptscriptstyle\bo}\dd{}_{{M}'})$
\end{description} 
\bea
&&{\rm Flat ~left}~\dron{\scriptscriptstyle\bo}\dd{}_{{M}}=
(S_{mn},~P_m,~\Sigma^{mn})~~;{\rm Flat ~right}~\dron{\scriptscriptstyle\bo}\dd{}_{{M}'}=(S_{m'n'}~P_{m'},~\Sigma^{m'n'})\nn\\
&&\left\{\begin{array}{ccl}
S_{mn}&=&\nabla_S\\
P_m&=&\nabla_P+\frac{1}{2}J_{1;P}\\
\Sigma^{mn}&=&\nabla_{\Sigma}+J_{1;\Sigma}
\end{array}\right.~~~~~~~~~
\left\{\begin{array}{ccl}
S_{m'n'}&=&\nabla_{S'}\\
P_{m'}&=&\nabla_{P'}-\frac{1}{2}J_{1;P'}\\
 \Sigma^{m'n'}&=&\nabla_{\Sigma'}-J_{1;\Sigma'}
\end{array}\right.\label{flatcovder}
\eea }
\item{
\begin{description}
  \item[
Flat symmetry generators: ] $
\dron{\scriptscriptstyle\bo}{\tilde{\dd}}{}_{\underline{M}}=\dron{\scriptscriptstyle\bo}{\tilde{\nabla}}{}_{\underline{M}}
+\displaystyle\frac{1}{2}\don{\scriptscriptstyle\bo}{\tilde{J}}_1{}^{\underline{L}}
(-\eta_{\underline{LM}}+\don{\scriptscriptstyle\bo}{\tilde{B}}{}_{\underline{LM}})=
(\dron{\scriptscriptstyle\bo}{\tilde{\dd}}{}_{{M}},\dron{\scriptscriptstyle\bo}{\tilde{\dd}}{}_{{M}'})$
\end{description}
\bea
&&{\rm Flat ~left}~\dron{\scriptscriptstyle\bo}{\tilde{\dd}}{}_{{M}}=
(\tilde{S}_{mn},~\tilde{P}_m,~\tilde{\Sigma}^{mn})~~~~~~~~\nn\\
&&\left\{\begin{array}{ccl}
\tilde{S}_{mn}&=&\tilde{\nabla}_S-
(\tilde{J}_{1;S}+c_S^P\tilde{J}_{1;P}
+c_S^{\Sigma}\tilde{J}_{1;\Sigma})
\\
\tilde{P}_m&=&\tilde{\nabla}_P
-\frac{1}{2}(\tilde{J}_{1;P}+c_P^\Sigma\tilde{J}_{1;\Sigma})
\\
\tilde{\Sigma}^{mn}&=&\tilde{\nabla}_{\Sigma}\end{array}\right.\nn\\\label{flatsymgen}\\
&&{\rm Flat ~right}~\dron{\scriptscriptstyle\bo}{\tilde{\dd}}{}_{{M}'}=(\tilde{S}_{m'n'}~\tilde{P}_{m'},~\tilde{\Sigma}^{m'n'})\nn\\
&&\left\{\begin{array}{ccl}
\tilde{S}_{m'n'}&=&\tilde{\nabla}_{S'}+
(\tilde{J}_{1;S'}+c_{S'}^{P'}\tilde{J}_{1;P'}+c_{S'}^{\Sigma'}\tilde{J}_{1;\Sigma'})
\\
\tilde{P}_{m'}&=&\tilde{\nabla}_{P'}+\frac{1}{2}(\tilde{J}_{1;P'}+c_{P'}^{\Sigma'}\tilde{J}_{1;\Sigma'})
\\
 \tilde{\Sigma}^{m'n'}&=&\tilde{\nabla}_{\Sigma'}\end{array}\right.\nn
\eea 
where coefficients $c_M^N$'s are given by $M_I{}^J$ determined from \bref{ddDJB} and  \bref{BtilB}.
Their explicit forms, in a particular parametrization, have been given in \cite{Hatsuda:2015cia}. }
\end{itemize}
The affine nondegenerate doubled Poincar\'{e} algebras generated by the covariant derivatives and 
the symmetry generators are given as:
\begin{itemize}
  \item{
\begin{description}
  \item[Affine flat algebra by covariant derivatives : $\dron{\scriptscriptstyle\bo}\dd{}_{\underline{M}}=(\dron{\scriptscriptstyle\bo}\dd{}_{{M}},\dron{\scriptscriptstyle\bo}\dd{}_{{M}'}
  )$
] 
\end{description}  
\bea
  \left\{\begin{array}{ccl}
\lbrack \dron{\scriptscriptstyle\bo}\dd{}_{M}(1),
\dron{\scriptscriptstyle\bo}\dd{}_{N}(2)]
&=&-i\don{\scriptscriptstyle\bo}{f}{}_{MN}{}^{L}
\dron{\scriptscriptstyle\bo}\dd{}_{L}\delta(2-1)
-i\eta_{MN}\partial_\sigma\delta(2-1)~~\\  
\lbrack \dron{\scriptscriptstyle\bo}\dd{}_{M'}(1),
\dron{\scriptscriptstyle\bo}\dd{}_{N'}(2)]
&=&-i\don{\scriptscriptstyle\bo}{f}{}_{M'N'}{}^{L'}
\dron{\scriptscriptstyle\bo}\dd{}_{L'}\delta(2-1)
-i\eta_{M'N'}\partial_\sigma\delta(2-1)~~\\
&=&i\don{\scriptscriptstyle\bo}{f}{}_{MN}{}^{L}
\dron{\scriptscriptstyle\bo}\dd{}_{L'}\delta(2-1)
+i\eta_{MN}\partial_\sigma\delta(2-1)~~~\label{flatcovMp}\\
\lbrack \dron{\scriptscriptstyle\bo}\dd{}_{M}(1),
\dron{\scriptscriptstyle\bo}\dd{}_{N'}(2)]
&=&0
\end{array}\right.~~~
\eea
}
\item{\begin{description}
  \item[Affine flat algebra by symmetry generators :]$\dron{\scriptscriptstyle\bo}{\tilde{\dd}}{}_{\underline{M}}=
  (\dron{\scriptscriptstyle\bo}{\tilde{\dd}}{}_{{M}},
  \dron{\scriptscriptstyle\bo}{\tilde{\dd}}{}_{{M}'}  )$
\end{description}
  \bea
  \left\{\begin{array}{ccl}
\lbrack \dron{\scriptscriptstyle\bo}{\tilde{\dd}}{}_{M}(1),
\dron{\scriptscriptstyle\bo}{\tilde{\dd}}{}_{N}(2)]
&=&i\don{\scriptscriptstyle\bo}{f}{}_{MN}{}^{L}
\dron{\scriptscriptstyle\bo}{\tilde{\dd}}{}_{L}\delta(2-1)
+i\eta_{MN}\partial_\sigma\delta(2-1)~~\\  
\lbrack \dron{\scriptscriptstyle\bo}{\tilde{\dd}}{}_{M'}(1),
\dron{\scriptscriptstyle\bo}{\tilde{\dd}}{}_{N'}(2)]
&=&i\don{\scriptscriptstyle\bo}{f}{}_{M'N'}{}^{L'}
\dron{\scriptscriptstyle\bo}{\tilde{\dd}}{}_{L'}\delta(2-1)
+i\eta_{M'N'}\partial_\sigma\delta(2-1)~~\\
&=&-i\don{\scriptscriptstyle\bo}{f}{}_{MN}{}^{L}
\dron{\scriptscriptstyle\bo}{\tilde{\dd}}{}_{L'}\delta(2-1)
-i\eta_{MN}\partial_\sigma\delta(2-1)~~~\label{flatsymMp}\\
\lbrack \dron{\scriptscriptstyle\bo}{\tilde{\dd}}{}_{M}(1),
\dron{\scriptscriptstyle\bo}{\tilde{\dd}}{}_{N'}(2)]
&=&0
\end{array}\right.~~~
\eea
}
\item{\begin{description}
  \item[Commutativity:] 
\end{description}
\bea
\lbrack \dron{\scriptscriptstyle\bo}{{\dd}}{}_{\underline{M}}(1),
\dron{\scriptscriptstyle\bo}{\tilde{\dd}}{}_{\underline{N}}(2)]
&=&0
\eea
}
\end{itemize}
The flat space is defined by 
the affine nondegenerate doubled Poincar\'{e} algebra
generated by the covariant derivative \bref{flatcovder}.
The symmetry generators \bref{flatsymgen} 
become physical symmetry generators and dimensional reduction constraints. 
\par
\vskip 6mm
\subsection{Dimensional reduction constraints and the section condition}

The symmetry generators obtained in \bref{flatsymgen}
satisfying in \bref{flatsymMp}
become
the physical total momentum and the physical  total Lorentz generators
\bea
\tilde{P}_{{\rm total};m}=\tilde{P}_m+ \tilde{P}_{n'}\delta^{n'}_m~~,~~
\tilde{S}_{{\rm total};mn}=\tilde{S}_{mn}-\tilde{S}_{m'n'}\delta^{m'}_m\delta^{n'}_n~~~,
\eea 
and 
 dimensional reduction constraints
\bea
&&\phi_m=\tilde{P}_m- \tilde{P}_{n'}\delta^{n'}_m=0\label{phiphi}\\
&&\lbrack \phi_m(1),\phi_n(2)]=-i(\tilde{\Sigma}_{mn}-\tilde{\Sigma}_{m'n'}\delta_m^{m'}
\delta_n^{n'})\delta(2-1)\nn\\
&&\Rightarrow~~\tilde{\Sigma}_{mn}=\tilde{\Sigma}_{m'n'}=0\nn~~~.
\eea

The worldsheet  $\tau$/$\sigma$-diffeomorphism
generators constructed with  the metrics in \bref{GGp} 
and the Virasoro algebra are given as
\bea
&&{\cal H}_\sigma=\frac{1}{2}
\dron{\scriptscriptstyle\bo}{{\dd}}{}_{\underline{M}}
\eta^{\underline{MN}}
\dron{\scriptscriptstyle\bo}{{\dd}}{}_{\underline{N}}~~,~~
{\cal H}_\tau=\frac{1}{2}
\dron{\scriptscriptstyle\bo}{{\dd}}{}_{\underline{M}}
\hat{\eta}^{\underline{MN}}
\dron{\scriptscriptstyle\bo}{{\dd}}{}_{\underline{M}}\\
&&{\renewcommand{\arraystretch}{1.4}
\left\{\begin{array}{ccl}
\lbrack {\cal H}_\sigma(1), {\cal H}_\sigma(2)]&=&i({\cal H}_\sigma(1)+
{\cal H}_\sigma(2))\partial_\sigma\delta(2-1)\\
\lbrack {\cal H}_\sigma(1), {\cal H}_\tau(2)]&=&i({\cal H}_\tau(1)+
{\cal H}_\tau(2))\partial_\sigma\delta(2-1)\\
\lbrack {\cal H}_\tau(1), {\cal H}_\tau(2)]&=&i({\cal H}_\sigma(1)+
{\cal H}_\sigma(2))\partial_\sigma\delta(2-1)
\end{array}\right.}~~~.\nn
\eea
These Virasoro constraints are imposed on the physical states for strings. 
This $\sigma$-diffeomorphism constraint written in the doubled coordinates 
is imposed  on the fields in the doubled target space, as the section condition.

The relation between the section condition
and the dimensional reduction constraint is the following:
The  $\sigma$-diffeomorphism constraint is satisfied on the constrained surface
\bea
{\cal H}_\sigma&=&
\frac{1}{2}(P_m{}^2-P_{m'}{}^2+\frac{1}{2}S_{mn}\Sigma^{mn}
-\frac{1}{2}S_{m'n'}\Sigma^{m'n'})\approx \frac{1}{2}
(P_m{}^2-P_{m'}{}^2)\nn\\
&=&\frac{1}{2}
\dron{\scriptscriptstyle\bo}{\tilde{\dd}}{}_{\underline{M}}
\eta^{\underline{MN}}
\dron{\scriptscriptstyle\bo}{\tilde{\dd}}{}_{\underline{N}}
\approx\tilde{P}_{{\rm total},m}\phi^m=0~~~
\eea
where  weak equalities $\approx$ in the first and the second lines 
are equal up to the  constraints, 
local Lorentz constraints $S_{mn}=S_{m'n'}=0$,
and the dimensional reduction constraints,  $\tilde{\Sigma}_{mn}=\tilde{\Sigma}_{m'n'}=0$.
In our formulation the first class constraint 
in \bref{phiphi} is imposed, which reduces to
$\frac{\partial}{\partial y_{\rm m}}\Phi=0$
only after the unitary gauge with $y_{\rm m}=x^m-x^{m'}$. 
The section condition is automatically satisfied.

The zero-modes of the symmetry generators
 satisfy the Poincar\'{e} algebra as
\bea
&&{\cal P}_{{\rm total};m}= \int d\sigma~\tilde{P}{}_{{\rm total};m}(\sigma)~,~
{\cal S}_{{\rm total};mn}~=~\int d\sigma~\tilde{S}{}_{{\rm total};mn}(\sigma)\nn\\
&&{\renewcommand{\arraystretch}{1.6}
\left\{\begin{array}{ccl}
\lbrack {\cal S}_{{\rm total};mn}, {\cal S}_{{\rm total};lk}]&=&
i\eta_{[k|[m}{\cal S}_{{\rm total};n]|l]}\\
\lbrack {\cal S}_{{\rm total};mn}, {\cal P}_{{\rm total};l}]&=&
i{\cal P}_{{\rm total};[m}\eta_{n]|l}\\
\displaystyle\lbrack {\cal P}_{{\rm total};m}, {\cal P}_{{\rm total};n}]
&=&i(\tilde{\Sigma}_{mn}-\tilde{\Sigma}_{m'n'}\delta_{m}^{m'}\delta_n^{n'})\approx 0
\end{array}\right.}\label{Flatalg}
~~~
\eea
where the dimensional reduction constraints \bref{phiphi} are used in the last equality.

\par\vskip 6mm
\section{Curved backgrounds in the asymptotically flat space}

\subsection{Curved space covariant derivative and torsion} 

The gravitational fields are coupled to closed string modes as 
given in \bref{orthogonal} 
\bea
\don{\scriptscriptstyle\bo}\dd_M~\to~\dd_{\underline{A}}=
E_{\underline{A}}{}^{\underline{M}}\don{\scriptscriptstyle\bo}\dd_M
\label{vieldd}~~,~~
E_{\underline{A}}{}^{\underline{M}}\eta_{\underline{MN}} 
E_{\underline{B}}{}^{\underline{N}}=\eta_{\underline{AB}} 
~~,~~E_{\underline{A}}{}^{\underline{M}}\eta^{\underline{AB}} 
E_{\underline{B}}{}^{\underline{N}}=\eta^{\underline{MN}} 
~~~.\label{OrthoE}
\eea
The vielbein fields
$E_{\underline{A}}{}^{\underline{M}}$
satisfies the  orthogonal condition with respect to $\eta_{\underline{MN}}$.
The generators includes Lorentz generators so the
vielbein includes not only $G_{mn}$ 
and $B_{mn}$  but also  the Lorentz connection $\omega_m{}^{nl}$
\cite{Polacek:2013nla}.

While the  $\sigma$-diffeomorphism generator in a curved space is unchanged from the one in a flat space because of the orthogonality 
\bref{OrthoE},
the $\tau$-diffeomorphism generator ${\cal H}_\tau$  in a curved space is given by
\bea
{\cal H}_\sigma&=&\frac{1}{2}\dd_{\underline{A}}\eta^{\underline{AB}}\dd_{\underline{B}}=\frac{1}{2}\don{\scriptscriptstyle\bo}\dd_{\underline{M}}\eta^{\underline{MN}}\don{\scriptscriptstyle\bo}\dd_{\underline{M}}
\nn\\
{\cal H}_\tau
&=&\frac{1}{2}\dd_{\underline{A}}\hat{\eta}^{\underline{AB}}\dd_{\underline{B}}
=\frac{1}{2}\don{\scriptscriptstyle\bo}\dd_{\underline{M}}{\cal M}^{\underline{MN}}
\don{\scriptscriptstyle\bo}\dd_{\underline{N}}~~,~~
{\cal M}^{\underline{MN}}=
E_{\underline{A}}{}^{\underline{M}}
\hat{\eta}^{\underline{AB}}
E_{\underline{B}}{}^{\underline{N}}~~
\eea
with the generalized metric ${\cal M}^{\underline{MN}}$ as a generalization of the third line of \bref{orthogonal}.
Since the  $\sigma$-diffeomorphism generator ${\cal H}_{\sigma}$
is independent on the background,
it is possible to impose  ${\cal H}_{\sigma}=0$
as a first class constraint even in curved spaces.

The  covariant derivative in a curved space $\dd_{\underline{A}}$
given in \bref{vieldd} satisfies the following algebra
\bea
&&
\lbrack {\dd}_{\underline{A}}(1),{\dd}_{\underline{B}}(2)]=-iT_{{\underline{AB}}}{}^{{\underline{C}}}{\dd}_{\underline{C}}\delta(2-1)
-i\eta_{{\underline{AB}}}\partial_\sigma\delta(2-1)~~~\nn\label{conaffalg}
\\
&&T_{\underline{ABC}}\equiv T_{\underline{AB}}{}^{\underline{D}}\eta_{\underline{DC}}
=\frac{1}{2}(i\nabla_{[\underline{A}}E_{\underline{B}}{}^{\underline{M}})E_{\underline{C}]\underline{M}}+
E_{\underline{A}}{}^{\underline{M}}E_{\underline{B}}{}^{\underline{N}}E_{\underline{C}}{}^{\underline{L}}\dron{\scriptscriptstyle\bo}f_{\underline{MNL}}~~~\label{torsion}\\
&&i\nabla_{[\underline{A}}T_{\underline{BCD}]}+\frac{3}{4}
T_{[\underline{AB}}{}^{\underline{E}}T_{\underline{CD}]\underline{E}}=0~~~.\nn
\eea
The  orthogonality of the vielbein is used 
to give the same Schwinger term as the flat case,
and the torsion 
$T_{\underline{AB}}{}^{\underline{D}}$ with lowered indices is totally antisymmetric.
The Bianchi identity leads to  the totally antisymmetric equation.

\par\vskip 6mm
\subsection{Group manifolds and the three form $H=dB$}

We focus on the cases where the curved space is a group manifold
so the torsion becomes  constant 
$T_{\underline{AB}}{}^{\underline{C}} \to 
{f}_{\underline{IJ}}{}^{\underline{K}}$. 
The covariant derivative 
${\dd}_{\underline{I}}=E_{\underline{I}}{}^{\underline{M}}
\don{\scriptscriptstyle\bo}\dd_{\underline{M}}$ 
satisfies the affine algebra as
\bea
\lbrack {\dd}_{\underline{I}}(1),{\dd}_{\underline{J}}(2)]=-i
{f}_{{\underline{IJ}}}{}^{{\underline{K}}}{\dd}_{\underline{K}}\delta(2-1)
-i\eta_{{\underline{IJ}}}\partial_\sigma\delta(2-1)~~~.\label{grmn}
\eea
From the expression of the torsion in \bref{conaffalg} 
the structure constant of the group manifold ${f}_{\underline{IJK}}$ 
is written in terms of vielbein field and 
the structure constant of the  nondegenerate doubled Poincar\'{e} algebra 
$ \dron{\scriptscriptstyle\bo}f_{\underline{MNL}}$ as
\bea
{f}_{\underline{IJK}}=\frac{i}{2}({\nabla}_{[\underline{I}}E_{\underline{J}}{}^{\underline{M}})E_{\underline{K}]}{}_{\underline{M}}
+E_{\underline{I}}{}^{\underline{M}}E_{\underline{J}}{}^{\underline{N}}
E_{\underline{K}}{}^{\underline{L}} \dron{\scriptscriptstyle\bo}f_{\underline{MNL}}~~~.\label{fmaru}
\eea 
The currents and the particle covariant derivatives are given by
\bea
&&
{\renewcommand{\arraystretch}{1.6}
\left\{\begin{array}{ccl}
\don{\scriptscriptstyle\bo}J_1{}^{\underline{L}}&=&\partial_\sigma Z^M
\don{\scriptscriptstyle\bo}{R}{}_{\underline{M}}{}^{\underline{L}}\\
{J}_1{}^{\underline{I}}&=&\partial_\sigma Z^M{R}_{\underline{M}}{}^{\underline{I}}
\end{array}\right.}~~,~~
{\renewcommand{\arraystretch}{1.6}
\left\{\begin{array}{ccl}
\dron{\scriptscriptstyle\bo}{\nabla}{}_{\underline{L}}&=&(\don{\scriptscriptstyle\bo}{R}{}^{-1})_{\underline{L}}{}^{\underline{M}}\frac{1}{i}\partial_{\underline{M}}\\
{\nabla}_{\underline{I}}&=&({R}{}^{-1})_{\underline{I}}{}^{\underline{M}}\frac{1}{i}\partial_{\underline{M}}
\end{array}\right.}~\nn\\
&&\dd_{\underline{I}}=E_{\underline{I}}{}^{\underline{L}}
\dron{\scriptscriptstyle\bo}\dd{}_{\underline{L}}~~\Rightarrow~~E_{\underline{I}}{}^{\underline{L}}=
({R}{}^{-1})_{\underline{I}}{}^{\underline{M}}
\don{\scriptscriptstyle\bo}{R}{}_{\underline{M}}{}^{\underline{L}}~~~.\label{Rdiamaru}
\eea
From equations in  \bref{Rdiamaru} 
the relation between 
structure constants in \bref{fmaru} becomes
\bea
{J}{}^{\underline{I}}\wedge{J}{}^{\underline{J}}
\wedge{J}{}^{\underline{K}}
{f}_{\underline{IJK}}+d{J}{}^{\underline{I}}\wedge{J}{}^{\underline{J}}
\eta_{\underline{IJ}}
=\don{\scriptscriptstyle\bo}J{}^{\underline{M}}\wedge \don{\scriptscriptstyle\bo}J{}^{\underline{N}}
\wedge \don{\scriptscriptstyle\bo}J{}^{\underline{L}}
\don{\scriptscriptstyle\bo}{f}{}_{\underline{MNL}}+d\don{\scriptscriptstyle\bo}J{}^{\underline{M}}\wedge \don{\scriptscriptstyle\bo}J{}^{\underline{N}}\eta_{\underline{MN}}~~~.\nn
\eea 
Using Maurer-Cartan equations in \bref{dJJJ}
the three forms in the doubled space in \bref{HH} are shown to be equal as
\bea
\don{\scriptscriptstyle\bo}{H}={H}~~,~~{\renewcommand{\arraystretch}{1.6}
\left\{\begin{array}{ccl}
\don{\scriptscriptstyle\bo}H&=&\frac{1}{3!}\don{\scriptscriptstyle\bo}J{}^{\underline{M}}\wedge \don{\scriptscriptstyle\bo}J{}^{\underline{N}}
\wedge \don{\scriptscriptstyle\bo}J{}^{\underline{L}}
\dron{\scriptscriptstyle\bo}{f}{}_{\underline{MNL}}\\
{H}&=&\frac{1}{3!}{J}{}^{\underline{I}}\wedge
{J}{}^{\underline{J}}
\wedge{J}{}^{\underline{K}}
{f}_{\underline{IJK}}
\end{array}\right.}\label{HmaruHdia}~~~.
\eea
In the doubled space the three form $H=dB$ is universal at least locally.
This is a consequence of the orthogonality in \bref{OrthoE} in the doubled formalism.

For the three form $\don{\scriptscriptstyle\bo}H={H}$ in 
\bref{HmaruHdia} the two form $B$ in the covariant derivatives in curved spaces $\dd_{\underline{A}}$
are equal up to the gauge  transformation
\bea
\don{\scriptscriptstyle\bo}B~~~~~~~~~~&=&~~~~~~~~~~~{B}+d\Lambda \nn\\
=\frac{1}{2}dZ^{\underline{M}}\wedge dZ^{\underline{N}}
(\don{\scriptscriptstyle\bo}{R}{}_{\underline{M}}{}^{\underline{L}}
\don{\scriptscriptstyle\bo}{R}{}_{\underline{N}}{}^{\underline{K}}
\don{\scriptscriptstyle\bo}{B}{}_{\underline{LK}})&=&\frac{1}{2}dZ^{\underline{M}}\wedge dZ^{\underline{N}}
({B}_{\underline{MN}}+\partial_{[\underline{M}}\Lambda_{\underline{N}]})
\\
=\frac{1}{2}
\don{\scriptscriptstyle\bo}{J}{}^{\underline{L}}\wedge \don{\scriptscriptstyle\bo}{J}{}^{\underline{K}}
\don{\scriptscriptstyle\bo}{B}{}_{\underline{LK}}~~
&=&\frac{1}{2}{J}{}^{\underline{I}}\wedge {J}{}^{\underline{J}}
\left({B}_{\underline{IJ}}+
({R}{}^{-1})_{\underline{I}}{}^{\underline{M}} 
({R}{}^{-1})_{\underline{J}}{}^{\underline{N}}
\partial_{[\underline{M}}\Lambda_{\underline{N}]}\right)\nn
\eea
$\don{\scriptscriptstyle\bo}{B}{}_{\underline{LK}}$
is the constant solution given in \bref{Bflatconstant}
and ${B}{}_{\underline{MN}}=R_{\underline{M}}{}^{\underline{I}}
R_{\underline{N}}{}^{\underline{J}}
{B}{}_{\underline{IJ}}$.
The $B$ field in the group manifold
and $\don{\scriptscriptstyle\bo}B$ field in the flat sapce 
are introduced in covariant derivatives as
\bea
{\renewcommand{\arraystretch}{1.6}
\left\{\begin{array}{ccl}
\dron{\scriptscriptstyle\bo}\dd{}_{\underline{M}}&=&\dron{\scriptscriptstyle\bo}{\nabla}{}_{\underline{M}}
+\displaystyle\frac{1}{2}\don{\scriptscriptstyle\bo}{J}_1{}^{\underline{N}}
(\eta_{\underline{NM}}+
\don{\scriptscriptstyle\bo}B{}_{\underline{NM}})
\\
{\dd}_{\underline{I}}&=&
{\nabla}_{\underline{I}}
+\displaystyle\frac{1}{2}{J}_1{}^{\underline{J}}(\eta_{\underline{JI}}+
{B}_{\underline{JI}})
\end{array}\right.}~~~
\eea
which  are related by the vielbein as in \bref{Rdiamaru}.
The gauge symmetry of ${B}_{\underline{MN}}$ field in the covaiant derivative 
${\dd}_{\underline{I}}$ is realized by the rotation between the momentum and the
winding mode as
\bea
\delta_\Lambda
{\dd}_{\underline{I}}&=&
({R}{}^{-1})_{\underline{I}}{}^{\underline{M}}~
\delta_\Lambda \left(\frac{1}{i}\partial_{\underline{M}}\right)+
\frac{1}{2}
({R}{}^{-1})_{\underline{I}}{}^{\underline{N}}
\partial_\sigma Z^{\underline{M}}
\partial_{[\underline{M}}\Lambda_{\underline{N}]}
\nn\\
&\Leftrightarrow&{\renewcommand{\arraystretch}{1.6}
\left(\begin{array}{c}
\frac{1}{i}\partial_{\underline{M}}\\
\partial_\sigma Z^{\underline{M}}
\end{array}
\right)\to
\left(\begin{array}{cc}
\delta_{\underline{M}}^{\underline{N}}
&\partial_{[\underline{N}}\Lambda_{\underline{M}]}
\\0&\delta^{\underline{M}}_{\underline{N}}
\end{array}
\right)
\left(\begin{array}{c}
\frac{1}{i}\partial_{\underline{N}}
\\
\partial_\sigma Z^{\underline{N}}
\end{array}
\right)}~~~.\label{T4B}
\eea
This transformation is a T-duality symmetry transformation
of the doubled momenta.
In other words the $B$ field in the doubled space 
is also recognized as a gauge field of 
the T-duality symmetry transformation given in \bref{T4B}.

\par
\vskip 6mm
\section{AdS space}
\subsection{Spontaneous symmetry breaking by the RR flux}

In this section we examine a bosonic  AdS space as a group manifold.
The AdS$_5\times$S$^5$ is a solution of the type IIB supergravity theory.
For the N=2 superalgebra
 the RR D3-brane charge  appears in
$\Upsilon_{\alpha\beta'}$ \cite{Hatsuda:2014aza}
\bea
\{D_\alpha,D_{\beta'}\}=\Upsilon_{\alpha\beta'}~~\label{DDUp}
\eea
where $D_\alpha$ and $D_{\beta'}$ are the left and right
supersymmetry charges
with $\alpha,\beta'=1,\cdots,16$.
On the other hand
the AdS$_5\times$S$^5$ 
superalgebra includes
the Lorentz terms $\frac{1}{r_{\rm AdS}}
(S\cdot \gamma)_{\alpha\beta}$
in the anticommutator of the  left and the right supercharges.  
The AdS$_5\times$S$^5$ space is obtained 
in the large D3-brane charge limit.
In the limit the right  hand side of \bref{DDUp} becomes
the product of the $1/r_{\rm AdS}$ times the Lorentz generator
$
\Upsilon_{\alpha\beta'}\to \frac{1}{r_{\rm AdS}}
(S\cdot \gamma)_{\alpha\beta}\epsilon_{12}$
with $S\cdot \gamma=S_{ab}\gamma^{ab}+S_{\bar{a}\bar{b}}\gamma^{\bar{a}\bar{b}}$,
 $a,b=0,1,\cdots,4$ and $\bar{a},\bar{b}=5,\cdots,9$,
and the vacuum expectation value of the RR flux becomes nonzero, 
$
\langle 0\mid F_{\rm RR}^{\alpha\beta'}\mid 0\rangle
=\frac{1}{r_{\rm AdS}}(\gamma_{01234}+\gamma_{56789}){}^{\alpha\beta'}
$.

For a bosonic algebra a left-right mixing term will be introduced 
 instead of the central extension of the superalgebra
\bref{DDUp} as 
\bea
\lbrack P_a,P_{b'}]=\Upsilon_{ab'}~~
.\label{vevUp}
\eea 
The existence of $f_{PP'}{}^{\Upsilon}$ and 
introducing the nondegenerate pair as
$\eta_{\Upsilon\digamma}={\bf 1}$ 
lead to the existence of $f_{PP'\digamma}$,
\bea
\lbrack \digamma^{ab'}, P_c]=\delta^a_cP^{b'}~~,~~\lbrack \digamma^{ab'}, P_{c'}]=-\delta^{b'}_{c'}P^{a}
.
\eea
This suggests that  $\Upsilon_{ab'}$ and $\digamma^{ab'}$ \cite{Hatsuda:2014aza}
correspond to 
the left-right mixing Lorenz generators, $\Sigma^{ab'}$ and $S_{ab'}$ respectively. 
The algebra is determined by
 the Jacobi identity.
 
As a result the number of generators of the doubled 10-dimensional flat space  
and the one for the  doubled AdS$_5\times$S$^5$ coincide as follows.
\bea
&
{\renewcommand{\arraystretch}{1.3}
\begin{array}{|c||c|c||c|c|}\hline
&{\rm Flat}&{\rm number}&{\rm AdS}_5\times{\rm S}^5&{\rm number}\\\hline\hline
{\rm Lorentz}&S_{mn}&45&S_{ab},~S_{\bar{a}\bar{b}}&10+10\\
&S_{m'n'}&45&S_{a'b'},~S_{\bar{a}'\bar{b}'}&10+10\\
&&&S_{ab'},~S_{\bar{a}\bar{b}'}&25+25\\
\hline
{\rm Momenta}&P_m&10&P_a,~P_{\bar{a}}&5+5\\
&P_{m'}&10&P_{a'},~P_{\bar{a}'}&5+5\\
\hline
{\rm Lorentz}&\Sigma^{mn}&45&\Sigma^{ab},~\Sigma^{\bar{a}\bar{b}}&10+10\\
{\rm nondegenerate}&\Sigma^{m'n'}&45&\Sigma^{a'b'},~\Sigma^{\bar{a}'\bar{b}'}&10+10\\
{\rm partner}&&&\Sigma^{ab'},~\Sigma^{\bar{a}\bar{b}'}&25+25\\
\hline
\end{array}}&~~
\eea
The indices in this table are the followings:
10-d.~flat indices are $m,m'=0,\cdots,9$,
AdS$_5$ indices are $a,a'=0,1,2,3,4$ and S$^5$ indices are $\bar{a},\bar{a}'=5,6,7,8,9$.
The subgroup H of the coset G/H is modified  by the spontaneous symmetry breaking. 
The subgroup is given schematically as follows;
\bea
&~~~~~~~~~{\rm left}~~~~~~~{\rm right}~~~~~~~~~~~~~
~~~~~~{\rm left}~~~~~~~{\rm right}~~~~~~~~~
~~~~~~~~~{\rm left}~~{\rm right}~~{\rm left}~~{\rm right}
&\nn\\
&~~~~~~~~~{\rm Poincar\acute{e}}~~~~~~~~~~~~~
~~~~{\rm AdS}~~~~{\rm S}~~~~{\rm AdS}~~~~~{\rm S}~
~~~~~~~~~~~~~~~~~~{\rm AdS}~~~~~~~~~~~~~{\rm S}~&\nn\\
&~~~~~~~~~{P_b}~~~~~~~~~{P_{b'}}~~~~~~~~~~~~~
~~~~{P_b}~~~~~{P_{\bar{b}}}~~~~~{P_{b'}}~~~~~{P_{\bar{b}'}}~
~~~~~~~~~~~~~{P_b}~~~~~{P_{{b}'}}~~~~~{P_{\bar{b}}}~~~~~{P_{\bar{b}'}}
&\nn\\
&{\renewcommand{\arraystretch}{2.8}
\begin{array}{c|c|c|}\cline{2-3}
{P_a}&~~S_{ab}~~&\\
\cline{2-3}{P_{a'}}&&~~S_{a'b'}~~\\\cline{2-3}
\end{array}}
\Rightarrow
{\renewcommand{\arraystretch}{1.5}
\begin{array}{c|c|c|c|c|}
\cline{2-5}{P_a}&S_{ab}&&S_{ab'}&\\
\cline{2-5}{P_{\bar{a}}}&~~~~&S_{\bar{a}\bar{b}}&&
S_{\bar{a}\bar{b}'}\\
\cline{2-5}{P_{{a}'}}&S_{a'b}&&S_{a'b'}&\\
\cline{2-5}{P_{\bar{a'}}}&&S_{\bar{a}'\bar{b}}
&&S_{\bar{a}'\bar{b}'}
\\\cline{2-5}
\end{array}}
=
{\renewcommand{\arraystretch}{1.5}
\begin{array}{c|c|c|c|c|}\cline{2-5}{P_a}&S_{ab}&S_{ab'}&&\\
\cline{2-5}{P_{a'}}&S_{a'b}&S_{a'b'}&&\\
\cline{2-5}{P_{\bar{a}}}&&~~~~&S_{\bar{a}\bar{b}}&
S_{\bar{a}\bar{b}'}\\
\cline{2-5}{P_{\bar{a'}}}&&&S_{\bar{a}'\bar{b}}&S_{\bar{a}'\bar{b}'}
\\\cline{2-5}
\end{array}}&\nn\\\label{SSppUp}
\eea
The number of degrees of freedom of $G_{mn}$ and $B_{mn}$
is ${\rm d}^2$ which now coincides with the number of the dimension 
of the coset  O(d,d)/
O($\frac{\rm d}{2},\frac{\rm d}{2}$)$^2$. 
 In this paper we focus on the doubled bosonic AdS part
of the AdS$^5\times$S$_5$ space
 which is  
the upper-left part of the third figure in  \bref{SSppUp}
 from now on.  The doubled bosonic sphere part of the  AdS$^5\times$S$_5$ space
is analyzed similarly   which is  
the lower-right part of the third figure in  \bref{SSppUp}.
\vskip 6mm

\subsection{Nondegenerate doubled AdS algebra}

At first  we make an AdS algebra doubled and nondegenerate
in this section.
In next subsection affine extension is performed.
The criteria of the  AdS algebra with manifest T-duality 
are followings:
\begin{itemize}
  \item{
Dimensional reduction of the doubled space algebra 
gives to the AdS algebra in the usual single coordinate space.
}
  \item{Doubled AdS algebra has a flat limit in the large AdS radius, $r_{\rm AdS}\to\infty$.}
  \item{Doubled AdS algebra has the nondegenerate group metric and the totally antisymmetric structure constant.}
  \end{itemize}

We focus on the bosonic 5-dimensional AdS part  in AdS$_5\times$S$^5$. 
As seen in the previous section 
the existence of the RR flux leads to the
 left/right mixing Lorentz generators.
The doubled d-dimensional AdS space is described 
by SO(d,d+1) group.
Next the nondegenerate pair of the Lorentz generators are introduced
by direct product of another Lorentz group SO(d,d).
The obtained group  SO(d,d+1)$\times$SO(d,d) is the 
doubled AdS algebra with the nondegenerate group metric and the totally antisymmetric structure constant.

\par\vskip 6mm
\subsubsection{Doubled AdS algebra }

We double the AdS group 
into the ones for left and right AdS groups, 
in addition to them we include the left/right mixing 
as seen in the previous section.
So the doubled AdS group will be SO(d,d+1).

The   doubled d-dimensional AdS algebra is given by so(d,d+1) 
generated by doubled momenta $p_{\underline{a}}=(p_a,~p_{a'})$,
doubled Lorentz $s_{\underline{ab}}=(s_{ab},~s_{a'b'};~s_{ab'})$
where 
$a$ and $a'$ runs 0 to d$-$1.
The proposed doubled algebra is
\bea
&
\lbrack G_A,G_B]=if_{AB}{}^CG_C~,~
\lbrack G_{A'},G_{B'}]=if_{A'B'}{}^{C'}G_{C'}~,~
\lbrack G_A,G_{B'}]=if_{AB'}{}^\Upsilon G_\Upsilon&\nn\\
&\lbrack G_\Upsilon,G_{A}]=if_{\Upsilon A}{}^{B'} G_{B'}~,~
\lbrack G_\Upsilon,G_{A'}]=if_{\Upsilon A'}{}^{B} G_{B}~&\nn\\
&\lbrack G_\Upsilon,G_{\Upsilon}]=if_{\Upsilon \Upsilon}{}^{A} G_{A}
+if_{\Upsilon \Upsilon}{}^{A'} G_{A'}
~~
&\label{GGmixed}
\eea
where the left/right mixed index is denoted by $\Upsilon$
including its nondegenerate partner $\digamma$ \cite{Hatsuda:2014aza}.
The doubled AdS algebra is given by 
 \bea
 {\renewcommand{\arraystretch}{1.6}
\begin{array}{lcl}
{\rm Left}&:&\lbrack s_{ab},s_{cd}]=i\eta_{[d|[a}s_{b]|c]}~,~
\lbrack s_{ab},p_{c}]=ip_{[a}\eta_{b]c}~,~
\lbrack p_{a},p_{b}]=i\frac{1}{r_{\rm AdS}{}^2}s_{ab}\\
{\rm Right}&:&\lbrack s_{a'b'},s_{c'd'}]=i\eta_{[d'|[a'}s_{b']|c'}~,~
\lbrack s_{a'b'},p_{c'}]=ip_{[a'}\eta_{b']c'}~,~
\lbrack p_{a'},p_{b'}]=i\frac{1}{r_{\rm AdS}{}^2}s_{a'b'}
\\
{\rm Mixed}&:&\lbrack s_{ab'},s_{cd'}]=
-i(\eta_{b'd'}s_{ac}+\eta_{ac}s_{b'd'})\\
&&\lbrack s_{ab},s_{cd'}]=
-i\eta_{c[a}s_{b]d'}
~,~\lbrack s_{a'b'},s_{cd'}]=
-i\eta_{d'[a'|}s_{c|b']}
\\&&\lbrack s_{ab'},p_{c}]=-i\eta_{ac}p_{b'}~,~
\lbrack s_{ab'},p_{c'}]=i\eta_{b'c'}p_{a}~,~
\lbrack p_{a},p_{b'}]=i\frac{1}{r_{\rm AdS}{}^2}s_{ab'}
\end{array}}\label{SOmix}
\eea
The spacetime metric of the  enlarged space is
\bea
 \eta_{\underline{ab}}=(\eta_{\natural\natural};\eta_{ab};\eta_{a'b'})
 =
 (-1;-1,1,1,1,1;1,-1,-1,-1,-1)~~~.
 \eea 
 The left moving mode is in an AdS space while
  the right moving mode is in a dS space. 
  This phenomena is similar to the point discussed in
\cite{Klimcik:2002zj}.
The structure constants with lowered indices 
$f_{\underline{ABC}}$ are totally antisymmetric. 

\par\vskip 6mm

\subsubsection{Nondegenerate doubled AdS algebra }

We will construct a nondegenerate AdS group 
SO(d,d+1)$\times$SO(d,d) in such a way that  the subalgebra H of the coset 
has its nondegenerate partner
by following the procedure given in subsection 2.2.1: 
\begin{enumerate}
\item{The doubled momenta are the generators of
the coset G/H$_0$ denoted by $ k$, w
here G=SO(d,d+1) is the doubled AdS group
and   H$_0$=SO(d,d) is the doubled Lorentz subgroups
and the left/right mixed Lorentz as in \bref{SOmix}.
}
\item{
Another Lorentz group H$_1$=SO(d,d) is introduced to construct 
the nondegenerate pair of the Lorentz group.
}
\item{Make nondegenerate pair $s_{\underline{ab}}$ and $\sigma^{\underline{ab}}$
by linear combinations of $h_0$ and $h_1$  which are 
Lie algebras of H$_0$ and H$_1$ as
\bea
&&{\renewcommand{\arraystretch}{1.6}\left\{\begin{array}{l}
h_0+ h_1=s\\
h_0-h_1=\displaystyle\frac{1}{r_{\rm AdS}{}^2}\sigma\\
k= \displaystyle\frac{1}{\sqrt{2}r_{\rm AdS}} p
\end{array}\right.}\nn\\
&&~\Rightarrow~
{\renewcommand{\arraystretch}{1.6}\left\{\begin{array}{l}
\lbrack s,s]=s~,~[s,\sigma]=\sigma~,~\lbrack \sigma,\sigma]=\displaystyle\frac{1}{r_{\rm AdS}{}^4}s\\
\lbrack s,p]=p~,~[p,p]=\displaystyle\frac{1}{r_{\rm AdS}{}^2}s+\sigma~,~[\sigma ,p]=\frac{1}{r_{\rm AdS}{}^2}p
\end{array}\right.}\label{ndAdS}
\eea
}
\item{Non-zero components of the 
nondegenerate doubled AdS group metrics are
\bea
\eta_{pp}=-\eta_{p'p'}={\bf 1}=
\eta_{s\sigma}=\eta_{s'\sigma'}=\eta_{\digamma\Upsilon}~~\eea
with $(p_a,p_{a'})=(p,p')$, $(s_{ab},s_{a'b'};s_{ab'})=(s,s';\digamma)$
and $(\sigma^{ab},\sigma^{a'b'};\sigma^{ab'})=(\sigma,\sigma';\Upsilon)$.
The signature of nondegenerate group metric  is determined from the Jacobi identity.
The  structure constant including constant torsions 
with lowered indices are totally antisymmetric:
\bea
&f_{ss\sigma}=-f_{s's'\sigma'}=
f_{\digamma\Upsilon s}=-f_{\digamma\Upsilon s'}=
f_{\digamma\digamma \sigma}=-f_{\digamma\digamma \sigma'}=
f_{pps}=f_{p'p's'}=-f_{pp'\digamma}={\bf 1}&\nn\\
&f_{pp\sigma}=f_{p'p'\sigma'}=
-f_{pp'\Upsilon}=\frac{1}{r_{\rm AdS}{}^2}{\bf 1}~~,~~
f_{\sigma\sigma\sigma}=-f_{\sigma'\sigma'\sigma'}
=f_{\Upsilon\Upsilon\sigma}=-f_{\Upsilon\Upsilon\sigma'}
=\frac{1}{r_{\rm AdS}{}^4}{\bf 1}~~~.&\nn\\
\eea
}
\end{enumerate}

\par
\vskip 6mm

\subsection{Affine AdS algebras}

\subsubsection{Covariant derivative and symmetry generator in the AdS space}

The covariant derivatives and the symmetry generators 
in the AdS space are given by
\bref{ddDJB} and \bref{symgenflat4}
as follows.
\begin{itemize}
\item{\begin{description}
  \item[AdS covariant derivatives] 
\end{description}
The covariant derivative in the AdS space is a linear combination of the 
AdS particle covariant derivative $\dron\circ{\nabla}_{\underline{A}}$
and the $\sigma$ component of the left-invariant current $\dron\circ{J}{}^{\underline{A}}$ with the $B$ field.
\bea
\dron\circ{\dd}_{\underline{A}}&=&
\don\circ{\nabla}_{\underline{A}}+\frac{1}{2}
\don\circ{J}{}^{\underline{B}}(\eta_{\underline{BA}}+
\don\circ{B}_{\underline{BA}})\label{AdScovder}
\eea
The $\don\circ{B}_{\underline{BA}}$ field on the AdS space
is a solution of the equation given in \bref{dBH}
and the existence of the solution is guaranteed by $d\don\circ{H}=0$.
The $B$ field on the AdS space  is not a constant
\bea
i\dron\circ{\nabla}_{[\underline{A}}\dron\circ{B}_{\underline{BC}]}-\dron\circ{f}_{[\underline{AB}|}{}^{\underline{D}}\dron\circ{B}_{\underline{D}|\underline{C}]} =
2\dron\circ{f}_{\underline{ABC}}~~~.
\eea
The covariant derivatives of the nondegenerate doubled AdS algebra
is the Lie algebra of the group SO(d,d+1)$\times$SO(d,d)
\bea
\dron\circ{\dd}_{\underline{A}}(\sigma)
&=&(S_{\underline{ab}},~P_{\underline{a}},~\Sigma^{\underline{ab}})~,~
{\renewcommand{\arraystretch}{1.4}\left\{\begin{array}{ccl}
S_{\underline{ab}}&=&(S_{ab},~S_{ab'},~S_{a'b'})\\
P_{\underline{a}}&=&(P_{a},~P_{a'})\\
\Sigma^{\underline{ab}}&=&(
\Sigma^{ab},~\Sigma^{ab'},~\Sigma^{a'b'})
\end{array}\right.}~~~.
\eea
}
\item{\begin{description}
  \item[AdS symmetry generators] 
\end{description}
The symmetry generator in the AdS space is a linear combination of the 
AdS particle symmetry generator $\dron\circ{\tilde{\nabla}}_{\underline{A}}$
and the $\sigma$ component of the right-invariant current $\dron\circ{\tilde{J}}{}^{\underline{A}}$ with the $\tilde{B}$ field.
\bea
\dron\circ{\tilde{\dd}}_{\underline{A}}&=&
\don\circ{\tilde{\nabla}}_{\underline{A}}+\frac{1}{2}
\don\circ{\tilde{J}}{}^{\underline{B}}(-\eta_{\underline{BA}}+
\don\circ{\tilde{B}}_{\underline{BA}})\label{AdSsymgen}\\
\don\circ{\tilde{B}}_{\underline{BA}}&=&
\don\circ{M}_{\underline{B}}{}^{\underline{C}}
\don\circ{M}_{\underline{A}}{}^{\underline{D}}
\don\circ{B}_{\underline{CD}}~,~
\don\circ{M}_{\underline{A}}{}^{\underline{D}}
=(\don\circ{L}{}^{-1})_{\underline{A}}{}^{\underline{M}}
\don\circ{R}_{\underline{M}}{}^{\underline{D}}~~~.\nn
\eea
The symmetry generators of the nondegenerate doubled AdS algebra
is the Lie algebra of the group SO(d,d+1)$\times$SO(d,d)
\bea
 \dron\circ{\tilde{\dd}}_{\underline{A}}(\sigma)
&=&(
S_{\underline{ab}},~\tilde{P}_{\underline{a}},~\tilde{\Sigma}^{\underline{ab}} )~,~
 {\renewcommand{\arraystretch}{1.4}\left\{\begin{array}{ccl}
 \tilde{S}_{\underline{ab}}&=&(\tilde{S}_{ab},~\tilde{S}_{ab'},~\tilde{S}_{a'b'})\\
 \tilde{P}_{\underline{a}}&=&
(\tilde{P}_{a},~\tilde{P}_{a'})\\\tilde{\Sigma}^{\underline{ab}}&=&(
\tilde{\Sigma}^{ab},~\tilde{\Sigma}^{ab'},~\tilde{\Sigma}^{a'b'})\end{array}\right.} ~~~.
\eea
}
\end{itemize}

\par
\vskip 6mm

\subsubsection{Affine AdS algebras}

The nondegenerate doubled AdS algebra in \bref{SOmix} and \bref{ndAdS}
is extended to affine AdS algebras generated by the AdS covariant 
derivative in \bref{AdScovder} and the AdS symmetry generator in \bref{AdSsymgen}. 
In contrast to the flat case
the left and right moving modes of the AdS algebra
are not really separated 
because of the left/right mixing caused by the RR flux.
Since the commutativity of the covariant derivative 
and the symmetry generator holds for the AdS space,
their roles hold in the AdS space; 
while the covariant derivative determine
the local structure of the space,
the symmetry generators are used to separate  out physical dimensions from unphysical dimensions.
The affine AdS algebras by the covariant derivative
\bref{AdScovder} and the symmetry generator \bref{AdSsymgen} in
components are listed as below.
\begin{itemize}
  \item{\begin{description}
  \item[Affine AdS algebras by  covariant derivative 
 $\dron\circ{\dd}_{\underline{A}}$ and symmetry generator 
 $\dron\circ{\tilde{\dd}}_{\underline{A}}$:] 
\end{description}
\bea
\lbrack \dron\circ{\dd}_{\underline{A}}(1),\dron\circ{\dd}_{\underline{B}}(2)]&=&-if_{{\underline{AB}}}{}^{{\underline{C}}}\dron\circ{\dd}_{\underline{C}}\delta(2-1)
-i\eta_{{\underline{AB}}}\partial_\sigma\delta(2-1)\nn\\
\lbrack  \dron\circ{\tilde{\dd}}_{\underline{A}}(1), \dron\circ{\tilde{\dd}}_{\underline{B}}(2)]&=&if_{{\underline{AB}}}{}^{{\underline{C}}} \dron\circ{\tilde{\dd}}_{\underline{C}}\delta(2-1)+i\eta_{{\underline{AB}}}\partial_\sigma\delta(2-1)\\
\lbrack \dron\circ{\dd}_{\underline{A}}(1), \dron\circ{\tilde{\dd}}_{\underline{B}}(2)]&=&0~~~\nn
\label{covsym}
\eea
}
\item{
\begin{description}
  \item[Affine AdS algebra by covaiant derivatives:] ~$\dron\circ{{\dd}}_{\underline{A}}=(\dron\circ{\dd}_A,~\dron\circ{\dd}_{A'},~\dron\circ{\dd}_{\Upsilon})$
\end{description}
\bea
&&{\rm AdS~Left:}~\dron\circ{\dd}_A=(S_{ab},~P_{a},~\Sigma^{ab})\nn\\
&&{\renewcommand{\arraystretch}{1.6}
\left\{\begin{array}{lcl}
 \lbrack S_{{ab}}(1),S_{{cd}}(2)]&=&{r_{\rm AdS}{}^4}\lbrack \Sigma_{{ab}}(1),\Sigma_{{cd}}(2)]~=~
 -i\eta_{[{d}|[{a}}S_{{b}]|{c}]}\delta(2-1)\\
\lbrack S_{{ab}}(1),P_{{c}}(2)]&=&{r_{\rm AdS}{}^2}\lbrack \Sigma_{{ab}}(1),P_{{c}}(2)]~=~-iP_{[{a}}\eta_{{b}]{c}}\delta(2-1)\\
\lbrack P_{{a}}(1),P_{{b}}(2)]&=&-i
(\displaystyle\frac{1}{r_{\rm AdS}{}^2}S_{{ab}}+\Sigma_{{ab}})\delta(2-1)
-i\eta_{{ab}}\partial_\sigma\delta(2-1)
\\
\lbrack S_{{ab}}(1),\Sigma_{{cd}}(2)]&=&-i
\eta_{[{d}|[{a}}\Sigma_{{b}]|{c}]}\delta(2-1)-i\eta_{{d}[{a}}\eta_{{b}]{c}}\partial_\sigma\delta(2-1)
\end{array}\right.}\label{result1}\\\nn\\
&&{\rm AdS~Right:}~\dron\circ{\dd}_{A'}=(S_{a'b'},~P_{a'},~\Sigma^{a'b'})\nn\\
&&{\renewcommand{\arraystretch}{1.6}\left\{\begin{array}{lcl}
 \lbrack S_{{a'b'}}(1),S_{{c'd'}}(2)]&=&{r_{\rm AdS}{}^4}\lbrack \Sigma_{{a'b'}}(1),\Sigma_{{c'd'}}(2)]~=~
-i\eta_{[{d'}|[{a}'}S_{{b}']|{c}']}\delta(2-1)\\
\lbrack S_{{a'b'}}(1),P_{{c}'}(2)]&=&{r_{\rm AdS}{}^2}\lbrack \Sigma_{{a'b'}}(1),P_{{c'}}(2)]~=~-iP_{[{a}'}\eta_{{b}']{c}'}\delta(2-1)\\
\lbrack P_{{a}'}(1),P_{{b}'}(2)]&=&
-i(\displaystyle\frac{1}{r_{\rm AdS}{}^2}S_{{a'b'}}+\Sigma_{{a'b'}})\delta(2-1)
-i\eta_{{a'b'}}\partial_\sigma\delta(2-1)
\\
\lbrack S_{{a'b'}}(1),\Sigma_{{c'd'}}(2)]&=&-i
\eta_{[{d'}|[{a'}}\Sigma_{{b'}]|{c'}]}\delta(2-1)-i\eta_{{d'}[{a'}}\eta_{{b'}]{c'}}\partial_\sigma\delta(2-1)
\end{array}\right.}\\\nn\\
&&{\rm AdS~Mixed:}~\dron\circ{\dd}_{\Upsilon}=(S_{ab'},~\Sigma^{{ab'}})
\nn\\
&&{\renewcommand{\arraystretch}{1.6}\left\{\begin{array}{lcl}
 \lbrack S_{ {ab'}}(1),S_{ {cd'}}(2)]&=&r_{\rm AdS}{}^4\lbrack \Sigma_{ {ab'}}(1),\Sigma_{ {cd'}}(2)]~=~
 i(\eta_{{b'}{d'}}S_{ac} +\eta_{ac}S_{b'd'})\delta(2-1)
 \\  \lbrack S_{ {ab}}(1),S_{ {cd'}}(2)]&=& r_{\rm AdS}{}^4 \lbrack \Sigma_{ {ab}}(1),\Sigma_{ {cd'}}(2)]~=~
 i\eta_{c[a}S_{b]d'}\delta(2-1)
 \\ 
 \lbrack S_{ {a'b'}}(1),S_{ {cd'}}(2)]&=& r_{\rm AdS}{}^4\lbrack \Sigma_{ {a'b'}}(1),\Sigma_{ {cd'}}(2)]~=~
i\eta_{{d'}[{a'}|}S_{c|b']}\delta(2-1)
 \\ 
\lbrack S_{ {ab'}}(1),P_{ {c}}(2)]&=&{r_{\rm AdS}{}^2}
\lbrack \Sigma_{ {ab'}}(1),P_{ {c}}(2)]~=~
i\eta_{ac}P_{b'}\delta(2-1)\\
\lbrack S_{ {ab'}}(1),P_{ {c'}}(2)]&=&
{r_{\rm AdS}{}^2}
\lbrack \Sigma_{ {ab'}}(1),P_{ {c'}}(2)]~=~
-i\eta_{b'c'}P_{a}\delta(2-1)\\
\lbrack P_{ {a}}(1),P_{ {b}'}(2)]&=&-i(\displaystyle\frac{1}{r_{\rm AdS}{}^2}S_{ {ab'}}+\Sigma_{ {ab'}})\delta(2-1)
\\
\lbrack S_{ {ab'}}(1),\Sigma_{ {cd'}}(2)]&=&i
(\eta_{{b}'d'}\Sigma_{ac}+\eta_{ac}\Sigma_{b'd'})
\delta(2-1)
+i\eta_{b'd'}\eta_{ac}\partial_\sigma\delta(2-1)
 \\  \lbrack S_{ {ab}}(1),\Sigma_{ {cd'}}(2)]&=&
 \lbrack \Sigma_{ {ab}} (1),S_{ {cd'}}(2)]~=~
 i\eta_{c[a}\Sigma_{b]d'}\delta(2-1)
 \\ 
 \lbrack S_{ {a'b'}}(1),\Sigma_{ {cd'}}(2)]&=&
 \lbrack \Sigma_{ {ab'}}(1),S_{ {cd'}}(2)]~=~
i\eta_{{d'}[{a'}|}\Sigma_{c|b']}\delta(2-1)
\end{array}\right.
}\label{2AdS}
\eea}

\item{
\begin{description}
  \item[Affine AdS algebra by symmetry generators:] ~$\dron\circ{\tilde{\dd}}_{\underline{A}}=(  \dron\circ{\tilde{\dd}}_A,~  \dron\circ{\tilde{\dd}}_{A'},~  \dron\circ{\tilde{\dd}}_{\Upsilon})$
\end{description}
\bea
&&{\rm AdS~Left:} ~ \dron\circ{\tilde{\dd}}_A=(\tilde{S}_{ab},~\tilde{P}_{a},~\tilde{\Sigma}^{ab})\nn\\
&&{\renewcommand{\arraystretch}{1.6}
\left\{\begin{array}{lcl}
 \lbrack \tilde{S}_{{ab}}(1),\tilde{S}_{{cd}}(2)]&=&{r_{\rm AdS}{}^4}
\lbrack \tilde{\Sigma}_{{ab}}(1),\tilde{\Sigma}_{{cd}}(2)]~=~
 i\eta_{[{d}|[{a}}\tilde{S}_{{b}]|{c}]}\delta(2-1)\\
\lbrack \tilde{S}_{{ab}}(1),\tilde{P}_{{c}}(2)]&=&{r_{\rm AdS}{}^2}\lbrack \tilde{\Sigma}_{{ab}}(1),\tilde{P}_{{c}}(2)]~=~i\tilde{P}_{[{a}}\eta_{{b}]{c}}\delta(2-1)\\
\lbrack \tilde{P}_{{a}}(1),\tilde{P}_{{b}}(2)]&=&i
(\displaystyle\frac{1}{r_{\rm AdS}{}^2}\tilde{S}_{{ab}}+\tilde{\Sigma}_{{ab}})\delta(2-1)
+i\eta_{{ab}}\partial_\sigma\delta(2-1)
\\
\lbrack \tilde{S}_{{ab}}(1),\tilde{\Sigma}_{{cd}}(2)]&=&i
\eta_{[{d}|[{a}}\tilde{\Sigma}_{{b}]|{c}]}\delta(2-1)+i\eta_{{d}[{a}}\eta_{{b}]{c}}\partial_\sigma\delta(2-1)
\end{array}\right.}\\\nn\\
&&{\rm AdS~Right:}~\dron\circ{\tilde{\dd}}_{A'}=(\tilde{S}_{a'b'},~\tilde{P}_{a'},~\tilde{\Sigma}^{a'b'})\nn\\
&&{\renewcommand{\arraystretch}{1.6}\left\{\begin{array}{lcl}
\lbrack \tilde{S}_{{a'b'}}(1),\tilde{S}_{{c'd'}}(2)]&=&
 {r_{\rm AdS}{}^4} \lbrack \tilde{\Sigma}_{{a'b'}}(1),\tilde{\Sigma}_{{c'd'}}(2)]~=~
i\eta_{[{d'}|[{a}'}\tilde{S}_{{b}']|{c}']}\delta(2-1)\\
\lbrack \tilde{S}_{{a'b'}}(1),\tilde{P}_{{c}'}(2)]&=&
 {r_{\rm AdS}{}^2}\lbrack \tilde{\Sigma}_{{a'b'}}(1),\tilde{P}_{{c'}}(2)]~=~
i\tilde{P}_{[{a}'}\eta_{{b}']{c}'}\delta(2-1)\\
\lbrack \tilde{P}_{{a}'}(1),\tilde{P}_{{b}'}(2)]&=&
i(\displaystyle\frac{1}{r_{\rm AdS}{}^2}\tilde{S}_{{a'b'}}+\tilde{\Sigma}_{{a'b'}})\delta(2-1)
+i\eta_{{a'b'}}\partial_\sigma\delta(2-1)
\\
\lbrack \tilde{S}_{{a'b'}}(1),\tilde{\Sigma}_{{c'd'}}(2)]&=&i
\eta_{[{d'}|[{a'}}\tilde{\Sigma}_{{b'}]|{c'}]}\delta(2-1)+i\eta_{{d'}[{a'}}\eta_{{b'}]{c'}}\partial_\sigma\delta(2-1)
\end{array}\right.}\\\nn\\
&&{\rm AdS~Mixed:}~\dron\circ{\tilde{\dd}}_{\Upsilon}=(\tilde{S}_{ab'},~\tilde{\Sigma}^{{ab'}})\nn\\
&&{\renewcommand{\arraystretch}{1.6}\left\{\begin{array}{lcl}
 \lbrack \tilde{S}_{ {ab'}}(1),\tilde{S}_{ {cd'}}(2)]&=&{r_{\rm AdS}{}^4}
\lbrack \tilde{\Sigma}_{ {ab'}}(1),\tilde{\Sigma}_{ {cd'}}(2)]~=~
  -i(\eta_{{b'}{d'}}\tilde{S}_{ac} +\eta_{ac}\tilde{S}_{b'd'})\delta(2-1)
 \\  \lbrack \tilde{S}_{ {ab}}(1),\tilde{S}_{ {cd'}}(2)]&=&{r_{\rm AdS}{}^4}  \lbrack \tilde{\Sigma}_{ {ab}}(1),\tilde{\Sigma}_{ {cd'}}(2)]~=~
 -i\eta_{c[a}\tilde{S}_{b]d'}\delta(2-1)
 \\ 
 \lbrack \tilde{S}_{ {a'b'}}(1),\tilde{S}_{ {cd'}}(2)]&=&{r_{\rm AdS}{}^4} \lbrack \tilde{\Sigma}_{ {a'b'}}(1),\tilde{\Sigma}_{ {cd'}}(2)]~=~
-i\eta_{{d'}[{a'}|}\tilde{S}_{c|b']}\delta(2-1)
 \\ 
\lbrack \tilde{S}_{ {ab'}}(1),\tilde{P}_{ {c}}(2)]&=&{r_{\rm AdS}{}^2}\lbrack \tilde{\Sigma}_{ {ab'}}(1),\tilde{P}_{ {c}}(2)]~=~-i\eta_{ac}\tilde{P}_{b'}\delta(2-1)\\
\lbrack \tilde{S}_{ {ab'}}(1),\tilde{P}_{ {c'}}(2)]&=&{r_{\rm AdS}{}^2}\lbrack \tilde{\Sigma}_{ {ab'}}(1),\tilde{P}_{ {c'}}(2)]~=~i\eta_{b'c'}\tilde{P}_{a}\delta(2-1)\\
\lbrack \tilde{P}_{ {a}}(1),\tilde{P}_{ {b}'}(2)]&=&
i(\displaystyle\frac{1}{r_{\rm AdS}{}^2}\tilde{S}_{ {ab'}}+\tilde{\Sigma}_{ {ab'}})\delta(2-1)
\\
\lbrack \tilde{S}_{ {ab'}}(1),\tilde{\Sigma}_{ {cd'}}(2)]&=&-i
(\eta_{{b}'d'}\tilde{\Sigma}_{ac}+\eta_{ac}\tilde{\Sigma}_{b'd'})
\delta(2-1)
-i\eta_{b'd'}\eta_{ac}\partial_\sigma\delta(2-1)
 \\  \lbrack \tilde{S}_{ {ab}}(1),\tilde{\Sigma}_{ {cd'}}(2)]&=&
 \lbrack \tilde{\Sigma}_{ {ab}}(1),\tilde{S}_{ {cd'}}(2)]~=~
- i\eta_{c[a}\tilde{\Sigma}_{b]d'}\delta(2-1)
 \\ 
 \lbrack \tilde{S}_{ {a'b'}}(1),\tilde{\Sigma}_{ {cd'}}(2)]&=&\lbrack \tilde{\Sigma}_{ {a'b'}}(1),\tilde{S}_{ {cd'}}(2)]~=~
-i\eta_{{d'}[{a'}|}\tilde{\Sigma}_{c|b']}\delta(2-1)
\end{array}\right.
}\label{2AdSsym}
\eea

}
\end{itemize}

\par
\vskip 6mm
\subsubsection{Curved backgrounds in the asymptotically AdS space}
 
The AdS space is spanned by the AdS covariant derivative $\dron\circ{\dd}
_{\underline{A}}$ in \bref{AdScovder}
which satisfies the
affine Lie algebra given in the first line of \bref{covsym}. 
Let us consider gravity theory as a fluctuation in 
the asymptotically AdS space as
\bea
\dd_{\underline{M}}=E_{\underline{M}}{}^{\underline{A}}\dron\circ{\dd}
_{\underline{A}}~~~.
\eea
The commutator of the covariant derivative gives the torsion and
the Bianchi identity gives the torsion equations
\bea
&&\lbrack \dd_{\underline{M}}(1),\dd_{\underline{N}}(2)]=-iT_{\underline{MNL}}\dd^{\underline{L}}
\delta(2-1)-i\eta_{\underline{MN}}\partial_\sigma\delta(2-1)\nn\\
&&T_{\underline{MNL}}=T_{\underline{MN}}{}^{\underline{K}}\eta_{\underline{KL}}
=\frac{1}{2}(i\nabla_{[\underline{M}}
E_{\underline{N}}{}^{\underline{A}})E_{\underline{L}]\underline{A}}
+E_{\underline{M}}{}^{\underline{A}}E_{\underline{N}}{}^{\underline{B}}
E_{\underline{L}}{}^{\underline{C}}
\don\circ{f}_{\underline{ABC}}~~~\\&&
i\nabla_{[\underline{M}}T_{\underline{NLK}]}
+\frac{3}{4}T_{[\underline{MN}}{}^{\underline{E}}T_{\underline{LK}]\underline{E}}=0~~~.\nn
\eea
The general gauge transformations are 
calculated from T-bracket given in \bref{Lam12} and
\bref{Tbracket}  by taking the vielbein field as
$\hat{\Lambda}_2=E_{\underline{M}}{}^{\underline{A}}\dron\circ{\dd}
_{\underline{A}}$,
 the gauge parameters as
  $\hat{\Lambda}_1=\int \Lambda^{\underline{A}}\dron\circ{\dd}
_{\underline{A}}$.
 The structure constant and the covariant derivative are specified as
  the AdS structure constant 
$\don\circ{f}_{\underline{AB}}{}^{\underline{C}}$
and the AdS covariant derivative $\dron\circ{\dd}_{\underline{A}}$. 
The vielbein field 
has gauge symmetries generated by the above bracket as
\bea
\delta_\Lambda E_{\underline{M}}{}^{\underline{A}}\dron\circ{\dd}_{\underline{A}}
&=&i\lbrack \hat{\Lambda}_1,
E_{\underline{M}}{}^{\underline{A}}\dron\circ{\dd}_{\underline{A}}]
\nn\\
(\delta_\Lambda E_{\underline{M}}{}^{\underline{A}})E_{\underline{N}\underline{A}}
&=&i\nabla_{[\underline{M}}(E_{\underline{N}]}{}^{\underline{A}}\Lambda_{\underline{A}})
-T_{\underline{MNL}}E^{\underline{LA}}\Lambda_{\underline{A}}~~~.
 \label{Tgaugetransf}
\eea
In the asymptotically flat limit the gauge symmetry transformation 
\bref{Tgaugetransf} is reduced to the one 
 with the structure constant of the nondegenerate Poincar\'{e} algebra.
 
\par
\vskip 6mm
\subsection{Auxiliary dimensions and physical dimensions  }

In order to  manifest T-duality symmetry 
we have enlarged the space not only by introducing the doubled coordinates
but also by introducing auxiliary dimensions  of the nondegeneracy. 
In this section
dimensional reduction constraints are obtained 
to reduce such unphysical dimensions.
We also construct the physical symmetry algebra
in terms of the symmetry generators written by doubled coordinates on the constrained surface.

\subsubsection{Dimensional reduction constraints }

As discussed in section 5.1 
the non-zero vacuum expectation value of the RR flux
in the AdS space,
 $\langle 0|F_{\rm RR}^{\alpha\beta'}|0\rangle\neq 0$, 
breaks two Lorentz symmetries preserving only  a combination of
the left and right Lorentz transformations as  
\bea
&&
\lbrack \frac{1}{2}\lambda^{ab}\tilde{S}_{ab}+ 
\frac{1}{2}\lambda^{a'b'}\tilde{S}_{a'b'},
\langle 0|F_{\rm RR}^{\alpha\beta'}|0\rangle]\nn\\
&&~~~~~=
\frac{1}{2}\lambda^{ab}(\gamma_{ab})^\alpha{}_\beta 
\langle 0|F_{\rm RR}^{\beta\beta'}|0\rangle
+\frac{1}{2}\lambda^{a'b'}
\langle 0|F_{\rm RR}^{\alpha\alpha'}|0\rangle
(\gamma_{a'b'}){}^{\beta'}{}_{\alpha'}~~~.\label{ssbmu}
\eea
In general $\langle 0|F_{\rm RR}^{\alpha\beta'}|0\rangle$
depends on the Lorentz coordinates, so it is transformed
under the Lorentz tranformations as above.
In a simple gauge where the left and right spinors are
the same chirality for the total Lorentz group, 
the vacuum expectation value of the five form RR flux is represented as  
$\langle 0| F_{\rm RR}^{\alpha\beta'}|0
\rangle=\frac{1}{r_{\rm AdS}}
\mu^{\alpha\beta'}$ with $\mu^{\alpha\beta'}=
\epsilon_{IJ}(\gamma_{01234}+\gamma_{56789})^{\alpha\beta}$
with $N=2$ spinor indices $I,J$.
Only one combination of the two Lorentz symmetries 
with parameters 
$\lambda_{ab}+\lambda_{a'b'}=0$
preserves the vacuum symmetry
from  $
\lbrack \gamma_{ab},\gamma_{01234}+\gamma_{56789}]=0
$.
Therefore the preserved Lorentz symmetry will be
$\tilde{S}_{ab}-\tilde{S}_{a'b'}$.
We introduce a parameter as a left-right mixing coefficient
defined by the vacuum expectation value of the following tensor
\bea
\langle 0|F^{\alpha\alpha'}_{\rm RR}F^{\beta\beta'}_{\rm RR}|0\rangle 
(\gamma_a)_{\alpha\beta}(\gamma^{b'})_{\alpha'\beta'}
 =\frac{{\rm tr}{\bf 1}}{r_{\rm AdS}{}^2}\chi_{a}{}^{b'}~~~.
\eea
It is possible to choose $\chi_{aa'}$  satisfies
\bea
\chi_{aa'}\chi_{bb'}\eta^{ab}=-\eta_{a'b'}~~,~~
\chi_{aa'}\chi_{bb'}\eta^{a'b'}=-\eta_{ab}~~~,\label{chichieta}
\eea
and it is inert under the Lorentz rotations,
for a case $\chi_{a}{}^{b'}=\delta_a^{b'}$. 

The criteria of the dimensional reduction constraints are the followings:
\begin{itemize}
  \item{ Constraints are written in terms of symmetry generators. The symmetry generators commute with the covariant derivatives, so the dimensional reduction constraints can reduce unphysical degrees of freedom without changing the local geometry.}
  \item{ The survived symmetry generated by the total momentum and the total Lorentz is the usual AdS algebra.}
  \end{itemize}
  
Before examining the dimensional~reduction constraints
we analyze the non-abelian doubled algebra.
If the doubled group is a direct product, generated by  $G$ and $G'$,
it has Z$_2$ structure
\bea
&\lbrack G,G]=G,~[G',G']=-G',~[G,G']=0&\nn\\
&\Theta_0=(G-G'),~\Theta_1=(G+G')&
\nn\\&\Rightarrow~
[\Theta_\mu,\Theta_\nu]=\delta_{\mu\nu}\Theta_0+\epsilon_{\mu\nu}\Theta_{\mu+\nu}=
\Theta_{\mu+\nu}~,~{\rm mod}~ 2,~_{\mu=(0,1)}~~.&\label{Z2}
\eea
However we have introduced the left-right mixed term  $\Upsilon$
as in \bref{SSppUp} and \bref{GGmixed},
then the Z$_2$ structure is generalized.
The antisymmetric and symmetric parts of $\Upsilon$ are denoted as
 $[\Upsilon]$ and $(\Upsilon)$.
The generalized Z$_2$ structure is given  as;
\bea
&\lbrack p,p]=s,~[p',p']=-s',~[p,p']=[\Upsilon]+(\Upsilon)&~\nn\\
&\Theta_0=(s-s'),~\Theta_1=[\Upsilon],~\Theta_{2}=(s+s'),~
\Theta_3=(\Upsilon)&\nn\\
&~\Rightarrow~~[\Theta_0,\Theta_0]=\Theta_{0}~,~[\Theta_0,\Theta_i]=\Theta_{i}~,~
[\Theta_i,\Theta_j]=\delta_{ij}\Theta_0+\epsilon_{ijk}\Theta_{k}~\label{genZ4}
,~_{i,j,k=(1,2,3)}~.&\label{modifiedZ4}
\eea
There are three sets of representations of the above algebra \bref{modifiedZ4}:
\begin{itemize}
\item
\begin{description}
  \item[Lorentz symmetry generator algebra with $\tilde{S}$] 
\end{description}
The linear combinations of the left and right 
Lorentz symmetry generators in \bref{2AdSsym} satisfy the above structure:
\bea
&&{\renewcommand{\arraystretch}{1.6}\begin{array}{ll}
\Theta_0=\tilde{S}_{ab}-\tilde{S}_{a'b'}\chi_a{}^{a'}\chi_b{}^{b'}~,~&
\Theta_1=\tilde{S}_{[a|b'|}\chi_{b]}{}^{b'}~~~\\
\Theta_2=\tilde{S}_{ab}+\tilde{S}_{a'b'}\chi_a{}^{a'}\chi_b{}^{b'}~,~&
\Theta_3=\tilde{S}_{(a|b'|}\chi_{b)}{}^{b'}
\end{array}}
\label{Z4sym}\\
&&
{\renewcommand{\arraystretch}{1.6}
\left\{\begin{array}{lcllcl}
\lbrack \Theta_0{}_{;ab},\Theta_{\mu}{}_{;cd}]&=&
-i\eta_{[c|[a}\Theta_{\mu;b]|d]},~_{\mu=0,1,2}~&,~~
\lbrack \Theta_0{}_{;ab},\Theta_{3}{}_{;cd}]&=&
-i\eta_{(c|[a}\Theta_{3;b]|d)}~\\
\lbrack \Theta_i{}_{;ab},\Theta_i{}_{;cd}]&=&
-i\eta_{[c|[a}\Theta_{0;b]|d]},~_{i=1,2}~&,~~
\lbrack \Theta_3{}_{;ab},\Theta_3{}_{;cd}]&=&
-i\eta_{(c|(a}\Theta_{0;b)|d)}~
\\
\lbrack \Theta_i{}_{;ab},\Theta_3{}_{;cd}]&=&
-i\eta_{(c|[a}\Theta_{3-i;b]|d)},~_{i=1,2}&,~~
\lbrack \Theta_2{}_{;ab},\Theta_1{}_{;cd}]&=&
-i\eta_{[c|[a}\Theta_{3;b]|d]}\end{array}\right.}\nn
\eea
where the worldvolume argument $\sigma$ is abbreviated.

  \item 
\begin{description}
  \item[Affine Lorentz symmetry generator algebra
  of subgroup H$_0$ with $\tilde{S}+\tilde{\Sigma}$] 
\end{description}
{\bea
&&{\renewcommand{\arraystretch}{1.6}\begin{array}{l}
 \check{\Theta}_0=(\tilde{S}_{ab}
 +{r_{\rm AdS}{}^2}\tilde{\Sigma}_{ab})
-(\tilde{S}_{a'b'}+{r_{\rm AdS}{}^2}\tilde{\Sigma}_{a'b'})\chi_a{}^{a'}\chi_b{}^{b'}\\
 \check{\Theta}_1=\tilde{S}_{[a|b'|}\chi_{b]}{}^{b'}+{r_{\rm AdS}{}^2}\tilde{\Sigma}_{[a|b'|}\chi_{b]}{}^{b'}~\\
 \check{\Theta}_2=(\tilde{S}_{ab}+{r_{\rm AdS}{}^2}\tilde{\Sigma}_{ab})
+(\tilde{S}_{a'b'}+{r_{\rm AdS}{}^2}\tilde{\Sigma}_{a'b'})
\chi_a{}^{a'}\chi_b{}^{b'}\\
 \check{\Theta}_3=\tilde{S}_{(a|b'|}\chi_{b)}{}^{b'}+{r_{\rm AdS}{}^2}\tilde{\Sigma}_{(a|b'|}\chi_{b)}{}^{b'}
\end{array}}\nn\\
\label{Z4symSigma}\\
&&{\renewcommand{\arraystretch}{1.6}\left\{\begin{array}{ccl}
\lbrack \check{\Theta}_0{}_{;ab}(1),\check{\Theta}_\mu{}_{;cd}(2)]&=&
-2i\eta_{[c|[a}\check{\Theta}_{\mu;b]|d]}\delta(2-1)-2i{r_{\rm AdS}{}^2}\delta_{\mu,0}\eta_{c[a}\eta_{b]d}
\partial_\sigma\delta(2-1)\nn\\
\lbrack \check{\Theta}_0{}_{;ab}(1),\check{\Theta}_{3}{}_{;cd}(2)]&=&
-2i\eta_{(c|[a}\check{\Theta}_{3;b]|d)}\delta(2-1)
\nn\\
\lbrack \check{\Theta}_i{}_{;ab}(1),\check{\Theta}_i{}_{;cd}(2)]&=&
-2i\eta_{[c|[a}\check{\Theta}_{0;b]|d]}\delta(2-1)
-2i{r_{\rm AdS}{}^2}\eta_{c[a}\eta_{b]d}\partial_\sigma\delta(2-1)\\
\lbrack \check{\Theta}_3{}_{;ab}(1),\check{\Theta}_3{}_{;cd}(2)]&=&
-2i\eta_{(c|(a}\check{\Theta}_{0;b)|d)}\delta(2-1)~
-2i{r_{\rm AdS}{}^2}\eta_{c(a}\eta_{b)d}\partial_\sigma\delta(2-1)
\\
\lbrack \check{\Theta}_i{}_{;ab}(1),\check{\Theta}_3{}_{;cd}(2)]&=&
-2i\eta_{(c|[a}\check{\Theta}_{3-i;b]|d)}\delta(2-1)\\
\lbrack \check{\Theta}_2{}_{;ab}(1),\check{\Theta}_1{}_{;cd}(2)]&=&
-2i\eta_{[c|[a}\check{\Theta}_{3;b]|d]}\delta(2-1)
\end{array}\right.}
\eea
with $~_{\mu=0,1,2}$ and $~_{i=1,2}~$.
}

 \item 
\begin{description}
  \item[Affine Lorentz symmetry generator algebra of subgroup
  H$_1$ with $\tilde{S}-\tilde{\Sigma}$]
\end{description}
{\bea
&&
{\renewcommand{\arraystretch}{1.6}\begin{array}{l}
 \check{\check{\Theta}}_0=(\tilde{S}_{ab}-{r_{\rm AdS}{}^2}\tilde{\Sigma}_{ab})
-(\tilde{S}_{a'b'}-{r_{\rm AdS}{}^2}\tilde{\Sigma}_{a'b'})\chi_a{}^{a'}\chi_b{}^{b'}\\
 \check{\check{\Theta}}_1=\tilde{S}_{[a|b'|}\chi_{b]}{}^{b'}-{r_{\rm AdS}{}^2}\tilde{\Sigma}_{[a|b'|}\chi_{b]}{}^{b'}~\\
 \check{\check{\Theta}}_2=(\tilde{S}_{ab}-{r_{\rm AdS}{}^2}\tilde{\Sigma}_{ab})
+(\tilde{S}_{a'b'}-{r_{\rm AdS}{}^2}\tilde{\Sigma}_{a'b'})
\chi_a{}^{a'}\chi_b{}^{b'}\\
 \check{\check{\Theta}}_3=\tilde{S}_{(a|b'|}\chi_{b)}{}^{b'}-{r_{\rm AdS}{}^2}\tilde{\Sigma}_{(a|b'|}\chi_{b)}{}^{b'}
\end{array}}\nn\\
\label{sZ4symSigma}
\\&&
{\renewcommand{\arraystretch}{1.6}\left\{\begin{array}{ccl}
\lbrack \check{\check{\Theta}}_0{}_{;ab}(1),\check{\check{\Theta}}_\mu{}_{;cd}(2)]&=&
-2i\eta_{[c|[a}\check{\check{\Theta}}_{\mu;b]|d]}\delta(2-1)+2i{r_{\rm AdS}{}^2}\delta_{\mu,0}\eta_{c[a}\eta_{b]d}
\partial_\sigma\delta(2-1)
\nn\\
\lbrack \check{\check{\Theta}}_0{}_{;ab}(1),\check{\check{\Theta}}_{3}{}_{;cd}(2)]&=&
-2i\eta_{(c|[a}\check{\check{\Theta}}_{3;b]|d)}\delta(2-1)
\nn\\
\lbrack \check{\check{\Theta}}_i{}_{;ab}(1),\check{\check{\Theta}}_i{}_{;cd}(2)]&=&
-2i\eta_{[c|[a}\check{\check{\Theta}}_{0;b]|d]}\delta(2-1)
+2i{r_{\rm AdS}{}^2}\eta_{c[a}\eta_{b]d}\partial_\sigma\delta(2-1)
\\
\lbrack \check{\check{\Theta}}_3{}_{;ab}(1),\check{\check{\Theta}}_3{}_{;cd}(2)]&=&
-2i\eta_{(c|(a}\check{\check{\Theta}}_{0;b)|d)}\delta(2-1)~
+2i{r_{\rm AdS}{}^2}\eta_{c(a}\eta_{b)d}\partial_\sigma\delta(2-1)
\\
\lbrack \check{\check{\Theta}}_i{}_{;ab}(1),\check{\check{\Theta}}_3{}_{;cd}(2)]&=&
-2i\eta_{(c|[a}\check{\check{\Theta}}_{3-i;b]|d)}\delta(2-1)\\
\lbrack \check{\check{\Theta}}_2{}_{;ab}(1),\check{\check{\Theta}}_1{}_{;cd}(2)]&=&
-2i\eta_{[c|[a}\check{\check{\Theta}}_{3;b]|d]}\delta(2-1)
\end{array}\right.}\nn
\eea
}
\end{itemize}

The  linear combinations of the doubled momenta 
\bea
\phi_{\pm;a}&=&\tilde{P}_{a}\pm\tilde{P}_{a'}\chi_{a}{}^{a'}\label{phiPP}~~\eea
 satisfy the following algebras with $\check{\Theta}_\mu$ and 
$\check{\check{\Theta}}_\mu$ 
\bea
{\renewcommand{\arraystretch}{1.6}
\begin{array}{ccl}
\lbrack \phi_{+;a}(1),\phi_{+;b}(2)]
&=&\frac{i}{r_{\rm AdS}{}^2}({\check{\Theta}}_{2;ab}+\check{\Theta}_{1;ab})
\delta(2-1)~~~
\\
\lbrack \phi_{-;a}(1),\phi_{-;b}(2)]
&=&\frac{i}{r_{\rm AdS}{}^2}(\check{\Theta}_{2;ab}-\check{\Theta}_{1;ab})
\delta(2-1)~~~
\\
\lbrack \phi_{+;a}(1),\phi_{-;b}(2)]
&=&
\frac{i}{r_{\rm AdS}{}^2}(\check{\Theta}_{0;ab}-\check{\Theta}_{3;ab})
\delta(2-1)+2i\eta_{ab}\partial_\sigma\delta(2-1)\\
\label{PPSgamma}
\lbrack \check{\Theta}_{0;ab}(1),\phi_{\pm;c}(2)]&=&
2i \phi_{\pm;[a}\eta_{b]c}\delta(2-1)\\
\lbrack \check{\Theta}_{i;ab}(1),\phi_{\pm;c}(2)]&=&
2i \phi_{\mp;[a}\eta_{b]c}\delta(2-1),~_{i=1,2}\\
\lbrack \check{\Theta}_{3;ab}(1),\phi_{\pm;c}(2)]&=&
2i \phi_{\pm;(a}\eta_{b)c}\delta(2-1)\\
\lbrack \check{\check{\Theta}}_{\mu;ab}(1),\phi_{\pm;c}(2)]&=&
\lbrack \check{\check{\Theta}}_{\mu;ab}(1),\check{\Theta}_{\nu;cd}(2)]=0
\end{array}}
\eea

We choose a set of first class constraints to reduce 
unphysical dimensions as
\bea
\phi_{-;a}&=&\tilde{P}_{a}-\tilde{P}_{a'}\chi_{a}{}^{a'}=0~\nn\\
\psi_{ab}&=&(\tilde{S}_{ab}+r_{\rm AdS}{}^2\tilde{\Sigma}_{ab})+(\tilde{S}_{a'b'}+r_{\rm AdS}{}^2\tilde{\Sigma}_{a'b'})\chi_{a}{}^{a'}\chi_{b}{}^{b'}
 -\tilde{S}_{[a|b'}\chi_{|b]}{}^{b'}-r_{\rm AdS}{}^2\tilde{\Sigma}_{[a|b'}\chi_{|b]}{}^{b'}\nn\\
&=&\check{\Theta}_{2;ab}-\check{\Theta}_{1;ab}=0~\label{con1}\\
\varphi_{ab}&=&(\tilde{S}_{ab}-r_{\rm AdS}{}^2\tilde{\Sigma}_{ab})
-(\tilde{S}_{a'b'}-r_{\rm AdS}{}^2\tilde{\Sigma}_{a'b'})\chi_{a}{}^{a'}\chi_{b}{}^{b'}
+\tilde{S}_{[a|b'}\chi_{|b]}{}^{b'}
-r_{\rm AdS}{}^2\tilde{\Sigma}_{[a|b'}\chi_{|b]}{}^{b'}\nn\\
&=&
\check{\check{\Theta}}_{0;ab}+\check{\check{\Theta}}_{1;ab}=0
~~~\nn
\eea
which satisfy the following algebra
\bea
\lbrack \phi_{-;a}(1),\phi_{-;b}(2)]&=&i\psi_{ab}\delta(2-1)\nn\\
\lbrack \varphi_{ab}(1),\varphi_{cd}(2)]
&=&4i\eta_{[d|[a}\varphi_{b]|c]}\delta(2-1)\\
{\rm others}&=&0~~~.\nn
\eea
The first class constraint
$\phi_{-;a}=0$ reduces the half of the degrees of feedom of doubled momenta.
We also impose the local Lorentz constraints
 $S_{\underline{ab}}=0$.
The first class constraints
 $\psi_{ab}=\varphi_{ab}=0$ can be imposed without conflicting with the local Lorentz constraints  by the same reason.

\par
\vskip 6mm
\subsubsection{Physical AdS algebra}
The physical global AdS algebra is constructed as follows.
We identify the total momentum and the total Lorentz generator as
\bea
\tilde{P}_{{\rm total};a}&=&
\frac{1}{2}(\tilde{P}_{a}+\tilde{P}_{a'}\chi_{a}{}^{a'})
+\frac{1}{2}\phi_{-;a}
=\frac{1}{2}(\phi_{+;a}+\phi_{-;a})=\tilde{P}_{a}\label{totalPS}
\\
\tilde{S}_{{\rm total};ab}
&=&\frac{1}{2}(\tilde{S}_{ab}-\tilde{S}_{a'b'}\chi_{a}{}^{a'}\chi_{b}{}^{b'}+\tilde{S}_{[a|b'}\chi_{b]}{}^{b'})+\frac{1}{4}(\psi_{ab}-\varphi_{ab})\nn\\
&=&\frac{1}{2}
(\Theta_{0;ab}+\Theta_{1;ab})
+\frac{1}{4}(\psi_{ab}-\varphi_{ab})
=\frac{1}{2}
(\tilde{S}_{ab}+r_{\rm AdS}{}^2 \tilde{\Sigma}_{ab})
~~~.\nn
\eea
The  total momentum and the total Lorentz symmetry generators 
in the flat space are the same as \bref{totalPS} with 
first class constarints,  
 $\phi_{-;a}
=\psi_{ab}=\varphi_{ab}=0$ and $\tilde{S}_{ab'}=0$. 
The physical global AdS algebra is generated 
by the zero mode of the total momenta and the total Lorentz
generator 
\bea
&&{\cal P}_{{\rm total};a}= \int d\sigma~\tilde{P}{}_{{\rm total};a}(\sigma)~,~
{\cal S}_{{\rm total};ab}~=~\int d\sigma~\tilde{S}{}_{{\rm total};ab}(\sigma)\nn\\
&&{\renewcommand{\arraystretch}{1.6}
\left\{\begin{array}{ccl}
\lbrack {\cal S}_{{\rm total};ab}, {\cal S}_{{\rm total};cd}]&=&
i\eta_{[d|[a}{\cal S}_{{\rm total};b]|c]}\\
\lbrack {\cal S}_{{\rm total};ab}, {\cal P}_{{\rm total};c}]&=&
i{\cal P}_{{\rm total};[a}\eta_{b]|c}\\
\displaystyle\lbrack {\cal P}_{{\rm total};a}, {\cal P}_{{\rm total};b}]
&=&i\frac{2}{r_{\rm AdS}{}^2}{\cal S}_{{\rm total};ab}
\end{array}\right.}\label{AdSalg}
~~~.
\eea
The doubled AdS momenta is not a simple  sum of the left and the right momenta, because of
 the left moving AdS momentum and the right moving dS momentum.
Although the physical global AdS spacetime generators  coincide with the left moving symmetry generators,
they are written in terms of  the doubled coordinates so
the T-duality symmetry is manifest.

The total dS algebra is obtained vice versa as follows:
The constraint $ \varphi_{ab}=0$ in \bref{con1} is 
  instead 
 \bea
 \varphi_{-;ab}
&=&
\check{\check{\Theta}}_{0;ab}-\check{\check{\Theta}}_{1;ab}=0
~~~\nn\\
\lbrack \varphi_{-;ab}(1),\varphi_{-;cd}(2)]
&=&4i\eta_{[d|[a|}\varphi_{-;|b]|c]}\delta(2-1)~~~.
\eea
The total dS momentum and Lornetz generators are 
\bea
\tilde{P}_{{\rm dS};a'}\chi_{a}{}^{a'}
&=&
\frac{1}{2}(\tilde{P}_{a}+\tilde{P}_{a'}\chi_{a}{}^{a'})
-\frac{1}{2}\phi_{-;a}
=\frac{1}{2}(\phi_{+;a}-\phi_{-;a})=\tilde{P}_{a'}\chi_{a}{}^{a'}\label{dS}
\\
\tilde{S}_{{\rm dS};a'b'}\chi_{a}{}^{a'}\chi_{b}{}^{b'}
&=&\frac{1}{2}
(\Theta_{0;ab}+\Theta_{1;ab})
-\frac{1}{4}(\psi_{ab}-\varphi_{-;ab})
=-\frac{1}{2}
(\tilde{S}_{a'b'}+r_{\rm AdS}{}^2 \tilde{\Sigma}_{a'b'})\chi_{a}{}^{a'}\chi_{b}{}^{b'}
~~~.\nn
\eea
The global  dS algebra is generated by 
\bea
&&{\cal P}_{{\rm dS};a'}= \int d\sigma~\tilde{P}{}_{{\rm dS};a'}(\sigma)~,~
{\cal S}_{{\rm dS};a'b'}~=~\int d\sigma~\tilde{S}{}_{{\rm dS};a'b'}(\sigma)\nn\\
&&{\renewcommand{\arraystretch}{1.6}
\left\{\begin{array}{ccl}
\lbrack {\cal S}_{{\rm dS};a'b'}, {\cal S}_{{\rm dS};c'd'}]&=&
i\eta_{[d'|[a'}{\cal S}_{{\rm dS};b']|c']}\\
\lbrack {\cal S}_{{\rm dS};a'b'}, {\cal P}_{{\rm dS};c'}]&=&
i{\cal P}_{{\rm dS};[a'}\eta_{b']|c'}\\
\displaystyle\lbrack {\cal P}_{{\rm dS};a'}, {\cal P}_{{\rm dS};b'}]
&=&-i\frac{2}{r_{\rm AdS}{}^2}{\cal S}_{{\rm dS};a'b'}
\end{array}\right.}\label{dSalg}
~~~.
\eea
Unphysical coordinates for doubled dimensions,
Lorentz and its nondegenerate partner can be gauged away
by using local symmetries generated by the first class constraints.
These first class constraints commute with the covariant derivatives,
so our dimensional reduction procedure preserves the T-duality gauge symmetry manifestly.

\par\vskip 6mm
\subsubsection{Comparison with the non-doubled AdS algebra}

We also mention the relation between the AdS algebra in this paper and  
our previous AdS algebra in \cite{Hatsuda:2001xf,Hatsuda:2014qqa}.
In the previous paper the AdS$_5\times$S$^5$ space is described by the 
PSU(2,2$|$4) coordinates.
A half of doubled coordinates 
are gauged away, and  only coordinates for the physical total momentum 
and the physical total Lorentz symmetry are used.
Gauge fixing conditions and 
corresponding first class constraints  
are given: 
\bea
{\rm Gauge~fixing~conditions}&&x^{m'}-x^m=v^{{mn}}=v^{{m'n'}}=
u^{mn'}=u^{m'n'}+u^{mn}=0\nn\\
{\rm First~class~constraints}&&
\phi_{-;a}=\psi_{ab}=\varphi_{ab}=S_{ab}=S_{a'b'}=S_{ab'}=0\nn\\
{\rm Second~class~constraints}&&v^{mn'}=\tilde{\Sigma}^{ab'}=0\label{fix}
\eea
After the gauge fixing  \bref{fix} the covariant derivatives become
as in \cite{Hatsuda:2014qqa}
\bea
\left\{\begin{array}{ccl}
P_a&=&\frac{1}{2}(\don\circ{\nabla}_P+J^P)\\
P_{a'}&=&\frac{1}{2}(\don\circ{\nabla}_P-J^P)\\
\end{array}\right.
\Rightarrow~\left\{\begin{array}{ccl}
[P_a,P_b]&=&\don\circ{\nabla}_S+J^S+\partial_\sigma\delta=
\dron\circ{\dd}_\Sigma+\partial_\sigma\delta\\
\lbrack P_a,P_{b'}]&=&\don\circ{\nabla}_S=\dron\circ{\dd}_{S;ab'}\\
\lbrack P_{a'},P_{b'}]&=&\don\circ{\nabla}_S-J^S-\partial_\sigma\delta=
\dron\circ{\dd}_{\Sigma'}-\partial_\sigma\delta
\end{array}\right.
\eea
In the right hand sides of the first and third lines of the algebras 
the particle component of the Lorentz covariant derivatives,
$\don\circ{\nabla}_S$, are identified with 
$\dron\circ{\dd}_\Sigma$ and 
$\dron\circ{\dd}_{\Sigma'}$, rather than $\dron\circ{\dd}_S$ and 
$\dron\circ{\dd}_{S'}$.
It is because $S$ and $S'$ satisfy the opposite sign structure constant
in the doubled AdS algebra \bref{SOmix},
so it cannot be equal consistently.
This is the same reason that the naive sum of momenta
$\tilde{P}+\tilde{P}'$ does not satisfy the AdS algebra globally in
\bref{AdSalg}. 
Lorentz generators are coset constraints $\dron\circ{\dd}_S=\dron\circ{\nabla}_S=0$,
so they are included in $\dron\circ{\dd}_\Sigma$'s.

In the gauge  \bref{fix} the covariant derivatives and the symmetry generators for 
the left and right moving modes  in the flat case as become 
\bea
{\renewcommand{\arraystretch}{1.6}
\begin{array}{cl}
{\rm Covariant~derivatives}:&
S=-S'=\Xi_u\frac{1}{i}\partial_u\\
&P=e^u\frac{1}{i}\partial_x +\frac{1}{2}e^{-u}\partial_\sigma x~,~
P'= e^u\frac{1}{i}\partial_x-\frac{1}{2}e^{-u}\partial_\sigma x\\
&\Sigma=\Sigma'=e^{-u}\partial_\sigma u=\Xi_u{}^{-1}\partial_\sigma u\\
{\rm Symmetry ~generators}:&
\tilde{S}=-\tilde{S}'=\Xi_{-u}\frac{1}{i}\partial_u+[x,\frac{1}{i}\partial_x]\\
&\tilde{P}=P' e^{-u}~,~
\tilde{P}'=P e^{-u}\\
&\tilde{\Sigma}=\tilde{\Sigma}'=0
\end{array}}\label{psu224}
\eea
Indices of generators and coordinates are abbreviated;
The order of contraction of the indices are also omitted as,
$e^u\frac{1}{i}\partial_x=(\partial_x)^{n} u_{nm}$.
The left-invariant one form gives
$e^{-iu\cdot s}de^{iu\cdot s}=\Xi_u^{-1}idu\cdot s$ with $u\cdot s=\frac{1}{2}
u^{mn}s_{mn}$.
The left and right modes of the Lorentz and $\Sigma$ generators are not independent respectively.
The left and right modes of the momentum symmetry generators are not independent from 
the right and  left modes of the momentum covariant derivatives.
In this gauge it is easy to see that
the commutator of the left and right  AdS momenta 
gives the Lorentz generator which is nonzero \cite{Hatsuda:2001xf,Hatsuda:2014qqa}. 
The supergroup 
 PSU(2,2$|$4) as the AdS$_5\times$S$^5$ group is a gauge fixed version of the
 fully manifestly  T-duality formulation.
 Both the left and right AdS$_5\times$S$^5$ groups 
 do not exist; only one kind of the momentum, Lorentz and no nondegenerate Lorentz partner exist.
Although the covariant derivative of SO(5,5)$\times$SO(5,5) exists as
in \bref{psu224}, it is not manifestly doubled AdS covariant.  
Furthermore the gauge invariant superstring action in the AdS space with manifestly T-duality requires  the formulation without the gauge fixing.

\par\vskip 6mm

\section{Summary}
We have proposed a manifestly T-dual formulation of group manifolds.
Especially the bosonic part of the AdS space with manifestly T-duality caused by the RR flux is examined in detail.
 It is a doubled AdS space which is defined by the affine nondegenerate doubled AdS algebra given by \bref{covsym}-\bref{2AdSsym}.
Contrast to the flat case the left and right momenta do not commute by the existence of
 the AdS curvature term. This mixing leads to that the left and right momenta are 
 in AdS and dS spaces respectively. We have found a set of first class constraints \bref{con1}
 to reduce unphysical dimensions of the doubled AdS space.
 This allows the manifest T-dual formulation without any gauge 
 fixing such as dual coordinate independence, $\partial^{\rm m}=0$, on doubled space functions.
 Then the physical AdS algebra is realized preserving whole doubled space coordinates 
 to manifest T-duality.

The $B$ field is extended in the nondegenerate doubled space,
and it gives the Wess-Zumino term for a string with the manifest T-duality.
 We have also shown that the doubled space three form $H=dB$ is at least locally universal for arbitrary group manifolds with the same dimension.
 
The obtained  doubled AdS space will be useful to clarify the structure of integrability and the
dual confromal symmetry which play important roles in AdS/CFT correspondence.
The supersymmetric extension and analysis of the ferminoic T-duality are 
important issues which are the future problems.
The doubled group manifold and AdS spaces will have many applications.

\par\vskip 6mm

\section*{Acknowledgements}
We thank  to  Martin Pol\'{a}{\v{c}}ek  for valuable discussions. 
M.H. would like to thank  Arkady Tseytlin, Ctirad Klimcik,  Jerzy Lukierski,
Yuho Sakatani, Kentaro Yoshida and  Tetsuji Kimura for fruitful discussions. 
She  acknowledges hospitality  of
the Simons Center for Geometry and Physics  during 
"Generalized Geometry \& T-dualities" in May, 2016 and
``the 2016 Summer Simons workshop in Mathematics and Physics" in July, 2016
where this work has been developed. 
She also  acknowledges hospitality  of
WG3 Meeting of COST Action MP 1405, Wroclaw, Poland
during "Noncommutative geometry, quantum symmetries and quantum gravity II, XXXVII Max Born Symposium" in July 2016.
 W.S. is
 supported in part by National Science Foundation Grant No. PHY-1316617.



\end{document}